\documentclass[10pt]{article}

\usepackage{xcolor}
\usepackage{amsmath}
\usepackage{float}
\usepackage{graphicx}
\usepackage{tabularx}
\usepackage[letterpaper, top = 1.0in, bottom = 1.0in, right=0.75in, left = 0.75in]{geometry}
\usepackage{amssymb} 
\usepackage{titling}
\usepackage{titlesec}
\usepackage{times}
\usepackage{ragged2e}
\usepackage{multirow}
\usepackage{blindtext}
\usepackage{graphics} 
\usepackage{epsfig} 
\usepackage{bm}
\usepackage{caption}
\usepackage{subcaption}
\usepackage[noadjust]{cite}
\usepackage[labelformat=simple]{subcaption}

\usepackage{dsfont}
\usepackage{relsize}
\usepackage{verbatim}

\date{}
\titleformat{\section}{\normalsize \bfseries \scshape}{\thesection}{1em}{}
\titleformat{\subsection}{\normalsize \bfseries}{\thesubsection}{1em}{}
\title{\Huge GPS Spoofing Mitigation and Timing Risk Analysis in Networked PMUs via Stochastic Reachability}

\author{Sriramya Bhamidipati,~\textit{University of Illinois at Urbana-Champaign} \\
	Grace Xingxin Gao,~\textit{Stanford University} }

\begin{document}
\maketitle
\thispagestyle{empty}

\section*{BIOGRAPHIES}
\noindent \textbf{Sriramya Bhamidipati} is a Ph.D. student in the Department of Aerospace Engineering at the University of Illinois at Urbana-Champaign, where she also received her master’s degree in 2017. She obtained her B.Tech.~in Aerospace from the Indian Institute of Technology, Bombay in 2015. Her research interests include GPS, power and control systems, artificial intelligence, computer vision, and unmanned aerial vehicles. 
~\\

\noindent \textbf{Grace Xingxin Gao} is an assistant professor in the Department of Aeronautics and Astronautics at Stanford University. Before joining Stanford University, she was an assistant professor at University of Illinois at Urbana-Champaign. She obtained her Ph.D. degree at Stanford University. Her research is on robust and secure positioning, navigation and timing with applications to manned and unmanned aerial vehicles, robotics, and power systems.

\section*{ABSTRACT}
To address PMU vulnerability against spoofing, we propose a set-valued state estimation technique known as Stochastic Reachability-based Distributed Kalman Filter (SR-DKF) that computes secure GPS timing across a network of receivers. Utilizing stochastic reachability, we estimate not only GPS time but also its stochastic reachable set, which is parameterized via probabilistic zonotope (p-Zonotope). While requiring known measurement error bounds in only non-spoofed conditions, we design a two-tier approach: We first perform measurement-level spoofing mitigation via deviation of measurement innovation from its expected p-Zonotope, and second perform state-level timing risk analysis via intersection probability of estimated p-Zonotope with an unsafe set that violates IEEE C37.118.1a-2014 standards. We validate the proposed SR-DKF by subjecting a simulated receiver network to coordinated signal-level spoofing. We demonstrate an improved GPS timing accuracy and successful spoofing mitigation via our SR-DKF. We validate the robustness of the estimated timing risk as the number of receivers are varied.

\section{INTRODUCTION} \label{sec:intro}
\begin{figure}[H]
	\setlength{\belowcaptionskip}{-4pt}
	\centering	\includegraphics[width=0.4\columnwidth]{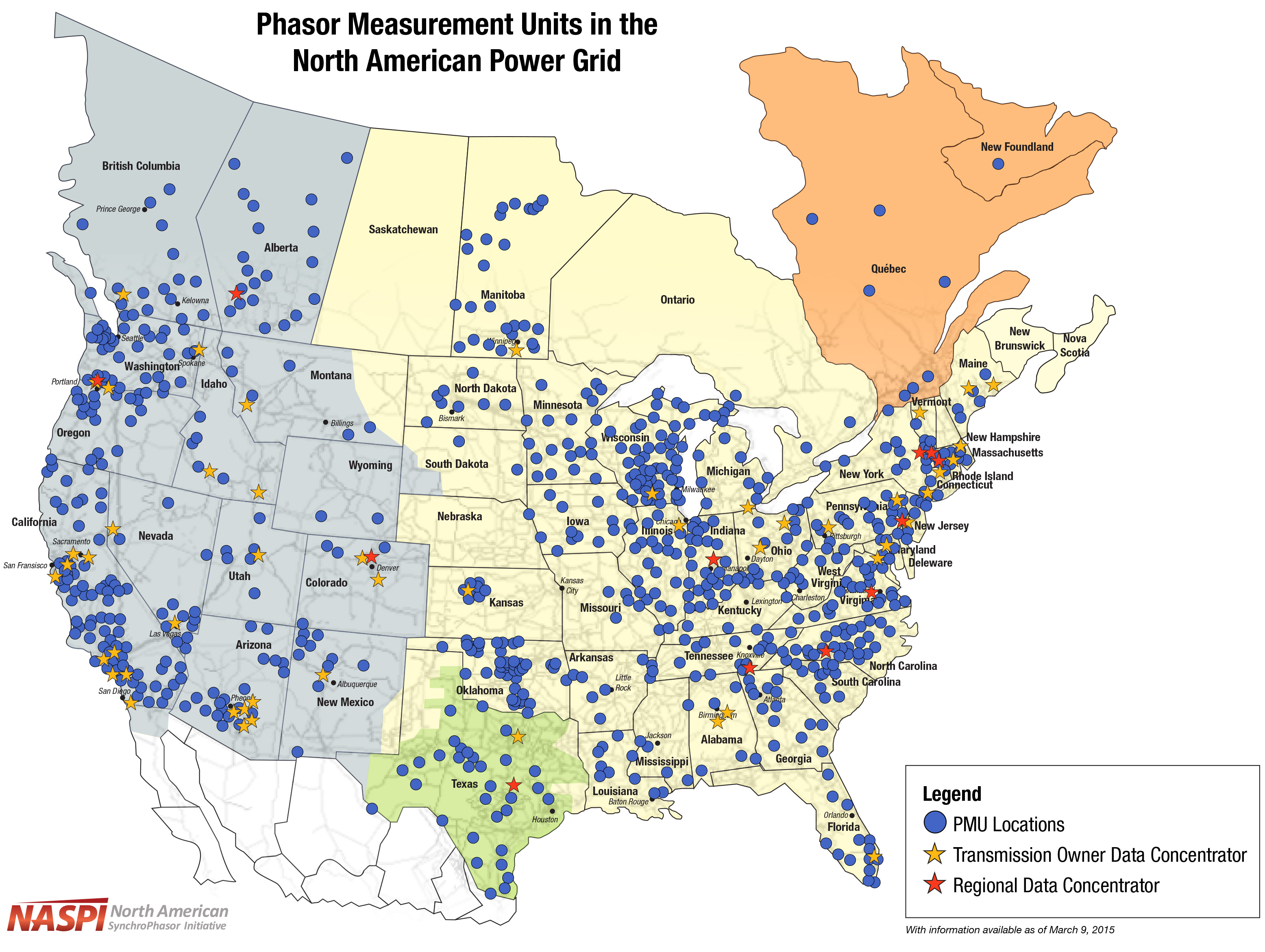}
	\caption{A network of $>2500$ PMUs deployed across the US~\cite{naspimap}}
	\label{fig_ION:synchrophasors}
\end{figure}

Modern power systems analyze real-time grid stability using a widely dispersed network of advanced devices, known as Phasor Measurement Units (PMUs)~\cite{duan2015development}. As per the American Recovery and Reinvestment Act of $2009$, almost $1700$ PMUs have been deployed across the US and Canada as of March $2015$~\cite{overholt2015synchrophasor,taft2018assessment}. 
Figure~\ref{fig_ION:synchrophasors} shows a map of the dense network of these spatially distributed PMUs. The North American Synchrophasor Initiative network (NASPInet), which is a collaborative effort by the U.S. Department of Energy and the North American Electric Reliability Corporation, provides a secure and decentralized infrastructure for communication among the PMUs~\cite{myrda2010naspinet}. Each PMU is equipped with a GPS receiver to precisely time-stamp the voltage and current phasor values at critical power substations. A network of geographically distributed GPS receivers facilitate robust time authentication by cross-checking each other and thereby eliminate the need for an external reference receiver~\cite{heng2014gps}. The networked approach in power systems overlays the existing grid infrastructure and therefore can be implemented without additional hardware. 

PMUs are susceptible to GPS spoofing because of the low signal power and unencrypted structure of the civilian GPS/L1 signals~\cite{misra2006global}. As validated in prior literature~\cite{shepard2012evaluation}, this vulnerability can be exploited by attackers to mislead the GPS timing at one or more critical substations and cause cascading system failure. The various spoofing attacks that can potentially disrupt GPS timing vary from a simple meaconing to more sophisticated signal-level spoofing and coordinated attacks. 
A spoofer executing meaconing (or time jump) broadcasts earlier recorded authentic GPS signals to induce a constant lag in the victim receiver’s computed time as compared to the true time. 
In signal-level spoofing (or time walk), the spoofer broadcasts the simulated GPS signals to gradually steer the computed time of victim receiver away from its true time. 
In addition, coordinated attack occurs when multiple spoofers manipulate the GPS signals received at one or more victim receivers.  
We refer readers to our prior work for a detailed overview of these attacks~\cite{bhamidipati2018gpsjournal}.

Secure GPS timing is of paramount importance for reliable PMU operations. The IEEE C37.118.1a-2014 standards for Synchrophasors~\cite{martin2015synchrophasor} define $1$\% Total Vector Error~(TVE) as a benchmark for grid stability. For a PMU operating at $60~$Hz, which is a commonly used frequency for phasor data logging, $>1$\% TVE is caused by a timing error of $>\pm26.5~\mu$s. The key measures of secure GPS timing are \textit{spoofing attack mitigation} and \textit{timing risk analysis}. In this context, we define risk as the probability of estimation error to exceed a safety Alert Limit~(AL). In this work, we set $AL=\pm26.5~\mu$s to ensure that the estimation error in GPS time complies with the IEEE C37.118.1a-2014 standards.  

\begin{figure}[h]
	\centering
	\begin{subfigure}[b]{0.35\textwidth}
		\includegraphics[width=\textwidth]{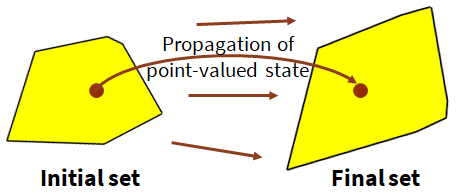}
		\caption{Set-valued approach}
		\label{fig:pointVsset}
	\end{subfigure}	
	\hspace{10mm}
	\begin{subfigure}[b]{0.5\textwidth}
		\includegraphics[width=\textwidth]{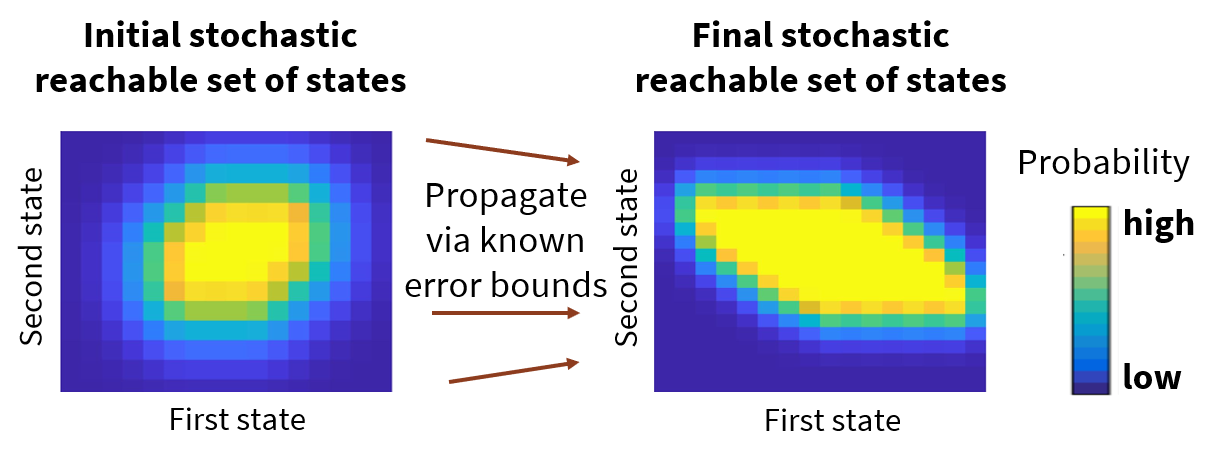}
		\caption{Stochastic reachability}
		\label{fig:sr_framework}
	\end{subfigure}
	\caption{Set-valued approach and stochastic reachability; (a)~shows an illustration of the difference between point-valued and set-valued approach. Unlike the point-valued techniques where only the point-valued state is propagated across time, the set-valued approach propagates the set of state estimates, thereby providing confidence bounds on the estimation errors; (b)~shows an intuitive understanding of stochastic reachability that computes a set of stochastic reachable states given an initial stochastic set and known error bounds. }
\end{figure}

The state-of-the-art prior work~\cite{yu2014short,khalajmehrabadi2018real,bhamidipati2018gps} on networked anti-spoofing falls under the category of point-valued state estimation approaches. A point-valued state estimate denotes the expected value of the state vector given the measurements. The existing literature on the point-valued anti-spoofing methods, which is summarized in~\cite{gunther2014survey,schmidt2016survey,bhamidipati2018gpsjournal}, includes signal quality monitoring, angle-of-arrival and time-difference-of-arrival-based analyses of the incoming satellite signals, validation of the encryption codes, and attack magnitude estimation. Unlike the point-valued techniques, a set-valued approach utilizes the pre-computed (or known) error bounds on system dynamics and received measurements to propagate the set of state estimates~\cite{shi2014set}. An illustration of the difference between point-valued and set-valued propagation is seen in Fig.~\ref{fig:pointVsset}. 

Much prior literature is available on set-valued theory~\cite{shi2014set,shiryaev2015set,combettes1993foundations}. Our current work is inspired by a set-valued formulation designed for stochastic processes known as stochastic reachability~\cite{prandini2006stochastic}. Stochastic reachability is widely used in robotics for path planning and collision avoidance, however, it has not been applied in the field of power systems. As shown in Fig.~\ref{fig:sr_framework}, stochastic reachability computes a set of stochastic reachable states given an initial stochastic set of states, and the known stochastic sets of system disturbances and measurement errors. In this context, each state in the stochastic reachable set is associated with a probabilistic measure that indicates the likelihood of its occurrence. Our choice of stochastic reachability is justified because it accounts for the stochastic errors in receiver clock dynamics, data transfer latencies, and GPS measurements. 

In this work, we propose a set-valued state estimation technique for a network of GPS receivers, termed as Stochastic Reachability-based Distributed Kalman Filter~(SR-DKF). The stochastic reachability formulation requires known error bounds, however, the attack errors induced via spoofing are unbounded and time-varying in nature. To account for this, we consider the error bounds in GPS measurements to be known only in non-spoofed (or authentic) conditions and pre-compute these bounds from offline empirical analysis of historical data~\cite{bhamidipati2018multiple}. This current work is based on our recent ION GNSS+ $2020$ conference paper~\cite{bhamidipati2020timing}.

In a conventional DKF approach~\cite{talebi2018distributed}, each receiver evaluates the received measurements against a point-valued attack threshold and thereafter fuses the measurements obtained from different receivers to compute its point-valued state estimate. In contrast, our SR-DKF performs secure state estimation of GPS time and spoofing attack mitigation in the set-valued domain. Therefore, we estimate not only the point-valued mean but also the stochastic reachable set of GPS time that is later utilized to compute the associated risk of violating IEEE C37.118.1a-2014 standards. Unlike the point-valued state estimation techniques~\cite{khalajmehrabadi2018real,bhamidipati2018gpsjournal}, our proposed SR-DKF algorithm does not require the conservative assumption that all satellite measurements at a victim receiver are attacked by the spoofer. Our set-valued analysis of spoofing attack combined with the redundancy in number of available measurements from a network of GPS receivers, eliminates the need to estimate the attack magnitude~\cite{alanwar2019distributed}. To our knowledge, no prior literature exists on performing the secure state estimation using stochastic reachability nor on quantifying the timing risk associated with estimated GPS time.

\subsection{Contributions}
The main contributions of this paper are listed as follows: 
\begin{enumerate}
	\item Utilizing an efficient set representation known as probabilistic zonotope~(p-Zonotope)~\cite{althoff2009safety} to parameterize the stochastic reachable set, we develop a set-valued DKF framework to compute secure GPS time for a network of receivers. 
	\item Considering offline-estimated measurement error bounds in non-spoofed (or authentic) conditions, we design a two-tier approach that performs
	\begin{enumerate}
		\item \textit{Measurement-level spoofing attack mitigation} by adaptively evaluating the deviation of point-valued GPS measurements from the set-valued domain of measurement innovation. Therefore, the proposed SR-DKF algorithm is resilient to a variety of GPS attacks, such as meaconing, signal-level spoofing, coordinated attack, etc. 
		\item \textit{State-level timing risk analysis} by estimating the intersection probability of the estimated set of stochastic reachable states (represented by p-Zonotope) with an unsafe set, i.e., a set violating the IEEE C37.118.1a-2014 standards~\cite{martin2015synchrophasor}. 
	\end{enumerate}
	\item For a simulated setup, we demonstrate the successful mitigation of both simple meaconing and sophisticated signal-level spoofing attacks. During a simulated case of coordinated attack that affects multiple receivers, we show an improvement in the state estimation accuracy via the proposed SR-DKF algorithm as compared to conventional point-valued DKF and single-receiver-based adaptive KF. We also demonstrate the robustness of the proposed SR-DKF algorithm by analyzing the variation in timing risk with respect to the number of receivers and magnitude of spoofing attacks.	
\end{enumerate}

The rest of the paper is organized as follows: Section~\ref{sec:problem} outlines the system model, the preliminaries of stochastic reachability and the conventional point-valued DKF; Section~\ref{sec:algorithm} describes the proposed SR-DKF algorithm; Section~\ref{sec:results} experimentally validates the improved state estimation accuracy of the GPS timing computed via SR-DKF as compared to the point-valued DKF, and also demonstrates the trend of timing risk associated with the estimated GPS time; and Section~\ref{sec:conclusion} concludes the paper.

\section{PROBLEM FORMULATION} \label{sec:problem}

The three main design aspects of the proposed SR-DKF algorithm, namely i) a spatially distributed network of static GPS receivers, ii) a distributed processing framework, and iii) the set representation via p-Zonotopes, are described as follows:

~\\
~\\
~\\
\noindent \textit{i)~A spatially distributed network of static GPS receivers:}

We utilize the existing grid infrastructure to consider a handful (around four to eight) of the spatially distributed PMUs (or GPS receivers). This framework provides two advantages: First, by considering a spatially distributed network, we increase the measurement redundancy while minimizing the probability of multiple receivers being simultaneously spoofed; Second, given that the power substations are static, we can utilize the known receiver positions to aid the proposed SR-DKF algorithm.

~\\
\noindent \textit{ii)~A distributed processing framework:} 

We opt for distributed processing to minimize the reaction time to attacks and also to localize the extent of failure. This is justified because global GPS reference time for synchronizing the PMUs is no longer reliable during spoofing. 
Therefore, the conventional communication protocol involving send or receive requests cannot be scheduled across the network. 
Each receiver in our distributed network asynchronously broadcasts the latest data to its neighbors, thereby reducing communication latency. 
\begin{figure}[h]
	\centering
	\begin{subfigure}[b]{0.35\textwidth}
		\includegraphics[width=\textwidth]{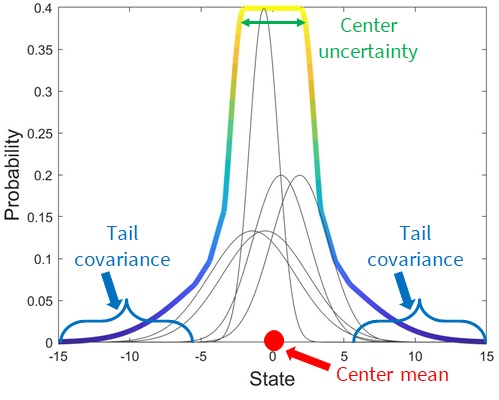}
		\caption{Example: 1D p-Zonotope}
		\label{fig:oneDimEx}
	\end{subfigure}
    \hspace{10mm} 
	\begin{subfigure}[b]{0.4\textwidth}
		\includegraphics[width=\textwidth]{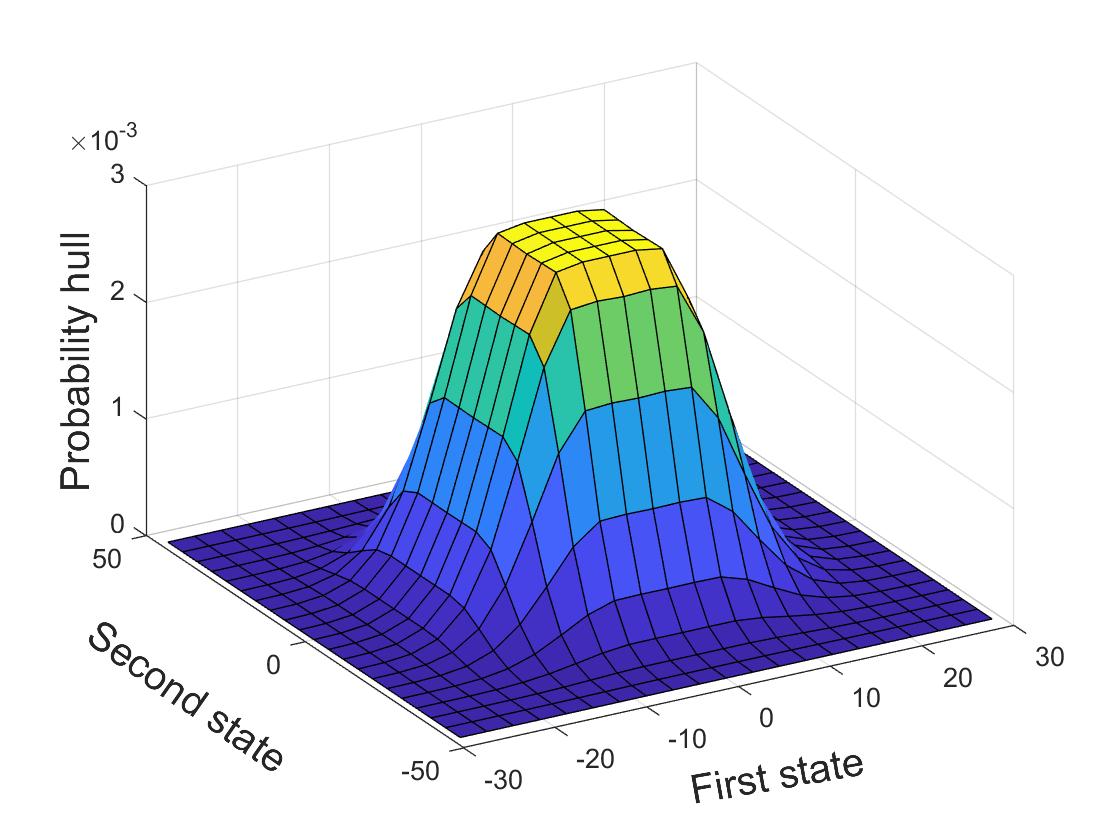}
		\caption{Example: 2D p-Zonotope}
		\label{fig:twodimEx}
	\end{subfigure}
	\caption{Illustration of p-Zonotopes; (a) shows the example of a 1D p-Zonotope and is indicated by the Parula colormap~\cite{nunez2018optimizing}. This 1D p-Zonotope encompasses multiple Gaussian distributions, a subset of which are shown in gray; (b) shows the example of a 2D p-Zonotope and is indicated by the Parula colormap. This 2D p-Zonotope is used to represent the state vector of interest to us in this paper and whose details are given later in Section~\ref{sec:algorithm}.}
\end{figure}

\noindent \textit{iii)~The set representation via p-Zonotopes:}

We represent stochastic reachable set via an enclosing probabilistic hull~\cite{althoff2009safety} known as p-Zonotope, denoted by $\mathcal{L}$. Among all available set representations~\cite{althoff2014online} such as zonotope, ellipsoid, polytope, etc., our choice of p-Zonotope is justified because it can efficiently enclose the stochastic biases in GPS measurements. A p-Zonotope can enclose a variety of distributions such as mixture models, distributions with uncertain/time-dependent mean, non-Gaussian/unknown distribution, etc. A one-Dimensional~(1D) example of a p-Zonotope that encloses multiple Gaussian distributions with varying mean and covariance is illustrated in Fig.~\ref{fig:oneDimEx}. The probabilistic nature of p-Zonotopes blends well with the Gaussian-centric framework of DKF.

Unlike the conventional techniques that propagate point-valued state estimates, the proposed SR-DKF algorithm utilizes the pre-computed error bounds on the receiver clock dynamics and measurements to recursively propagate the p-Zonotope of the state estimate. An example of a two-Dimensional~(2D) p-Zonotope is shown in Fig.~\ref{fig:twodimEx}. In this work, the 2D p-Zonotope is used to represent the state estimate that is introduced later in Section~\ref{sec:algorithm}. 

\subsection{Preliminaries of p-Zonotopes} \label{sec:pZonotopes}
A p-Zonotope $\mathcal{L}$, as seen in Eq.~\eqref{eq:pZono} and Fig.~\ref{fig:oneDimEx}, is characterized by
three parameters~\cite{althoff2009safety}, namely, a)~the mean of its center, denoted by $\bm{c}\in \mathbb{R}^{n}$, b)~the uncertainty in the center, represented by the generator matrix~$G\in \mathbb{R}^{n\times e}$, and c)~the over-bounding covariance of the Gaussian tails, denoted by $\Sigma\in \mathbb{R}^{n\times n}$. 
\begin{subequations}
	\begin{align} \label{eq:pZono}
	\mathcal{L}=(\bm{c},G,\Sigma)
	\end{align}
	Intuitively, a p-Zonotope with a certain center mean, i.e., zero center uncertainty~$G=\bm{0}$, represents a Gaussian distribution that overbounds multiple distributions with the same center mean~$\bm{c}$. This is mathematically represented by $\mathcal{Z}$ in Eq.~\eqref{eq:Gzono}. 
	\begin{align} \label{eq:Gzono}
	\mathcal{Z} = \Bigg\{\bm{c}+\sum_{i=1}^{l}\mathcal{N}^{i}(0,1)\,\underline{\bm{g}}^{i}\Bigg\},
	\end{align}
	\noindent where $\mathcal{N}^{i}(0,1)~\forall i\in \{1,\cdots,l\}$ denotes independent normally distributed random variables and $\underline{\bm{g}}^{i}\in \mathbb{R}^{n}$ denotes the associated generator vectors of~$\mathcal{Z}$, such that $\underline{G}=[\underline{\bm{g}}^{1},\cdots,\underline{\bm{g}}^{l}]$ and $\Sigma=\underline{G}\underline{G}^{\top}$. 
	
	Similarly, a p-Zonotope with zero covariance~$\Sigma=\bm{0}$ simplifies to a zonotope~\cite{althoff2014online} that encompasses all the possible values of the center, such that the generator matrix of~$Z$ is $G=[\bm{g}^{1},\cdots,\bm{g}^{e}]$. This case is mathematically represented by $Z$ in Eq.~\eqref{eq:zono}. 
	\begin{align} \label{eq:zono}
	Z = \Big\langle\bm{c}, G \Big\rangle=\Bigg\{\bm{c}+\sum_{i=1}^{e}\beta_{i}\,\bm{g}^{i}\Bigg| -1\leq \beta_{i}\leq 1 \Bigg\}
	\end{align}
\end{subequations} 

The p-Zonotope is a combination of both sets, i.e., $\mathcal{L}=\mathcal{Z}\boxplus Z$, where $\boxplus$ is a set-addition operator, such that \newline $\mathcal{L}=sup \big\{ f_{\mathcal{Z}} \Big| \mathbb{E}[\mathcal{Z}]\in Z\big\}$, where $f_{\mathcal{Z}}$ is the Probability Density Function~(PDF) of the distributions represented by $\mathcal{Z}$ and $\mathbb{E}[\cdot]$~denotes the expectation operator. From Eqs.~\eqref{eq:pZono}-\eqref{eq:zono}, note that  $\mathcal{L}$ depends on both $G$ and $\underline{G}$. Additionally, note that unlike a PDF, the area enclosed by a p-Zonotope does not equal to one. More details regarding the characteristics of zonotopes and p-Zonotopes can be found in the prior literature~\cite{althoff2009safety,alanwar2019distributed}. 

For example, the 2D p-Zonotope shown in Fig.~\ref{fig:twodimEx} is formulated by considering the bounds on the mean and covariance of the first state to be $[-5,5]$ and $[0,2]$, respectively, and of the second state to be $[-10,10]$ and $[0,3]$, respectively. Mathematically, we represent this 2D p-Zonotope~$\mathcal{L}=(\bm{c}, G, \Sigma)$ by $\bm{c}=\begin{bmatrix} 0 \\ 0 \end{bmatrix}$, $G=\begin{bmatrix} 
2 & 0 \\
0 & 3 
\end{bmatrix}$ and $\Sigma=3\times\begin{bmatrix} 
2 & 0 \\
0 & 3 
\end{bmatrix}=
\begin{bmatrix} 
6 & 0 \\
0 & 9
\end{bmatrix}$. We consider a multiplication factor~$3$ in the over-bounding covariance~$\Sigma$ based on the empirical rule according to which $99.7\%$ of the values lie within three standard deviations of the mean~\cite{grafarend2006linear}.

~\\
To perform set operations on the stochastic reachable sets, we require three important properties~\cite{althoff2009safety}:  
\begin{itemize}
	\item \textit{Minkowski sum of addition: } The addition of two p-Zonotopes $\mathcal{L}_{1}=(\bm{c}_{1}, G_{1}, \Sigma_{1})$ and $\mathcal{L}_{2}=(\bm{c}_{2}, G_{2}, \Sigma_{2})$ is given by
	\begin{subequations} 
		\begin{align} \label{eq:pZonoProp1}
		\mathcal{L}_{1}\bigoplus\mathcal{L}_{2} &= (\bm{c}_{1}+\bm{c}_{2},[G_{1},G_{2}],\Sigma_{1}+\Sigma_{2}),
		\end{align}
		\begin{align} \label{eq:pZonoProp1B}
		\bigoplus_{i=1}^{p}\mathcal{L}_{i} &= (\sum_{i=1}^{p}\bm{c}_{i},[G_{1},\cdots,G_{p}],\sum_{i=1}^{p}\Sigma_{i}),
		\end{align}
		\noindent where~$\bigoplus$ is known as the Minkowski sum operator and $[\cdot,\cdot]$ denotes a horizontal concatenation operator.
		\item \textit{Linear map: } Given a matrix $A\in \mathbb{R}^{n\times n}$ and a p-Zonotope $\mathcal{L}=(\bm{c},G,\Sigma)$, 
		\begin{align} \label{eq:pZonoProp2}
		A\mathcal{L}&=(A\bm{c},AG,A\Sigma A^{\top})
		\end{align}
		\item \textit{Translation: } Given a vector $\mu\in \mathbb{R}^{n}$ and a p-Zonotope $\mathcal{L}=(\bm{c},G,\Sigma)$, 
		\begin{align} \label{eq:pZonoProp3}
		\mu+\mathcal{L}&=(\mu+\bm{c},G,\Sigma)
		\end{align}
	\end{subequations}
\end{itemize}

\subsection{System model} \label{sec:prelim_system_model}
Based on the key features of the SR-DKF algorithm, we mathematically represent the inter-connected network of power substations, each mounted with one GPS receiver, as a undirected graph $\mathcal{G}=\{M,E\}$, where $M$ indicates the set of nodes and $E$ denotes the set of connections between the nodes. Each node represents a GPS receiver in the network and each edge indicates the presence of a secure bidirectional communication link between the power substations. The total number of receivers in the network are $\big|M\big|$, where $\big|\cdot\big|$ denotes the cardinality of a set. Our SR-DKF algorithm does not require the network of power substations to be fully connected, and hence, $\big|E\big|\leq \dfrac{\big|M\big|\big(\big|M\big|-1\big)}{2}$.  

The neighborhood set of any $i$th receiver~$\forall i\in M$ is represented by $M_{i}$ and indicates the subset of nodes among the $M$ nodes that can communicate with the $i$th receiver, including itself. Therefore, the associated cardinality of the neighborhood set is denoted by $\big|M_{i}\big|$ where $\big|M_{i}\big|\geq 1~\forall i\in M$. Every receiver in the network has a prior knowledge regarding the pre-computed static position of all the receivers in its neighborhood set. 

We define the global state vector at the $k$th time iteration, which is represented by $\bm{x}_{k}\in\mathbb{R}^{2}$, to be comprised of GPS time and its drift rate, which are given by $T_{k}$ and $\dot{T}_{k}$, respectively. As explained earlier in Section~\ref{sec:intro}, estimated GPS time~$T_{k}$ is used in PMUs for time-tagging the phasor measurements. At the $k$th iteration, the global state vector is the same across all the receivers in the network. Additionally, we define the local state vector at the $i$th receiver by $\bm{y}_{i,k}\in\mathbb{R}^{2}$. The local state vector comprises two states, namely, the receiver clock bias and clock drift, which are denoted by $c\delta t_{k}$ and $c\delta \dot{t}_{k}$, respectively. At the $i$th receiver, the global and local state vectors are related via secure time-varying local clock parameters, i.e., $T^{i}_{rx,k}$ and $\Delta T^{i}_{rx,k}$, which denote the measured local clock time and the sampling time, respectively. Therefore, $\bm{x}_{k}=\bm{y}_{i,k}+\bm{\theta}_{i,k}$ where $\bm{\theta}_{i,k}=[T^{i}_{rx,k},\Delta T^{i}_{rx,k}]^{\top}$. At the $k$th iteration, each $i$th receiver maintains an estimate of both the local and global state vectors. When a receiver is spoofed, the local state vector~$\bm{y}_{i,k}$ is manipulated, thereby misleading the global GPS time~$T_{k}$. 
 
The state-space model representing the true system at the $i$th receiver is given by 
\begin{align}
\begin{aligned}
\bm{x}_{k} &= F\bm{x}_{k-1}+\bm{\nu}_{i,k}, \\
\bm{z}_{i,k} &= H_{i}\bm{x}_{k}+\bm{\omega}_{i,k},~~~~~\forall i\in M,
\end{aligned}
\end{align}
\noindent where 
\begin{itemize}
	\item [--] $\bm{z}_{i,k}$ denotes the $2N_{i}\times 1$ translated GPS measurement vector, such that $\bm{z}_{i,k}=\Big[\tilde{\rho}^{1}_{i,k},\tilde{\phi}^{1}_{i,k},\cdots,\tilde{\rho}^{N_{i}}_{i,k},\tilde{\phi}^{N_{i}}_{i,k}\Big]$, where $N_{i}$ denotes the number of visible GPS satellites from the $i$th receiver. Also, $\tilde{\rho}^{j}_{i,k}$ and $\tilde{\phi}^{j}_{i,k}$ denote the GPS pseudorange and Doppler residual associated with the $j$th satellite~$\forall j\in\{1,\cdots,N_{i}\}$, respectively, such that \newline
	$\begin{bmatrix} 
	\tilde{\rho}^{j}_{i,k} \\ 
	\tilde{\phi}^{j}_{i,k}
	\end{bmatrix}=
	\begin{bmatrix} 
	\rho^{j}_{i,k} \\ 
	\phi^{j}_{i,k}
	\end{bmatrix}+\bm{\theta}_{i,k}-
	\begin{bmatrix} 
	||\bm{p}_{i}-\bm{p}^{j}|| \\ 
	||\dot{\bm{p}}_{i}-\dot{\bm{p}}^{j}||
	\end{bmatrix}$. Here, $\rho^{j}_{i,k}$ and $\phi^{j}_{i,k}$ denote the measured GPS pseudorange and Doppler value. Also, $\bm{p}_{i}$ and $\dot{\bm{p}}_{i}$ denote the known position and zero velocity of the $i$th static receiver, respectively, and $\bm{p}^{j}$ and $\dot{\bm{p}}^{j}$ denote the pre-computed position and velocity of the $j$th satellite, respectively;
	\item [--] $F$ denotes the first-order linear state transition model, such that $F = \begin{bmatrix} 1 & \Delta T^{i}_{rx,k} \\ 0 & 1
	\end{bmatrix}$, $f_{L1}$ is the frequency of the GPS L1 signal, and $c$ is the speed of light; 
	\item [--]  $H_{i}$ denotes the GPS measurement model, such that $H_{i}=\bm{1}_{2N_{i}\times 2}$ where $\bm{1}_{a\times b}$~denotes the matrix of ones with $a$~rows and $b$~columns;
	\item [--] $\bm{\nu}_{i,k}$ denotes the associated $2\times 1$ process noise vector and $\bm{\omega}_{i,k}$ denotes the $2N_{i}\times 1$ measurement noise vector that is associated with the GPS pseudoranges and Doppler values. 
\end{itemize}

\subsection{Preliminaries of the point-valued DKF } \label{sec:pntDKF}
Based on~\cite{talebi2018distributed}, the general framework of a point-valued DKF implementation performed independently at the $i$th receiver, is shown in Eqs.~\eqref{eq:pDKF_init}-\eqref{eq:pDKF_meas}. During initialization, i.e., $k<0$, we define the predicted value of the point-valued mean and covariance that are given by $\hat{\bm{x}}_{i,-1}$ and $\hat{P}_{i,-1}$, respectively. The process and measurement noise are modeled via a zero-mean Gaussian distribution that are given by $\nu_{i,k}\sim\mathcal{N}(\bm{0},Q_{i,k})$ and $\omega_{i,k}\sim\mathcal{N}(\bm{0},R_{i,k})$, respectively, where $Q_{i,k}$ and $R_{i,k}$ represent the process and measurement noise covariance matrix, respectively. After initialization, the conventional point-valued DKF recursively executes two steps: one is the measurement update and the other is the time update. 
\begin{align} \label{eq:pDKF_init}
\begin{split}
\hat{\bm{x}}_{i,-1} &= \mathbb{E}\big[\bm{x}_{0}\big]  \\
\hat{P}_{i,-1} &= \mathbb{E}\Big[\Big(\bm{x}_{0}-\mathbb{E}\big[\bm{x}_{0}\big]\Big)\Big(\bm{x}_{0}-\mathbb{E}\big[\bm{x}_{0}\big]\Big)^{\top}\Big]  \\
\end{split}
\end{align}

During the time update, as seen in Eq.~\eqref{eq:pDKF_time}, the corrected variables $\bar{\bm{x}}_{i,k-1}$ and $\bar{P}_{i,k-1}$ from the previous time, i.e., at the $(k-1)$th iteration, are propagated forward-in-time to predict the estimates for the next $k$th iteration, which are given by $\hat{\bm{x}}_{i,k}$ and $\hat{P}_{i,k}$, respectively. 
\begin{align} \label{eq:pDKF_time}
\begin{split}
\hat{\bm{x}}_{i,k} &= F\bar{\bm{x}}_{i,k-1} \\
\hat{P}_{i,k} &= F\bar{P}_{i,k-1}F^{\top}+Q_{i,k}
\end{split}
\end{align}

At the $k$th iteration, each $i$th receiver receives the GPS measurements~$\bm{z}_{j,k}$ and measurement noise covariance matrix~$R_{j,k}$ from the other receivers in the neighborhood set, i.e.,~$j\in M_{i}$, and processes them sequentially during the measurement update. In the measurement update, the expected measurements (obtained via the predicted estimates) and the received measurements are weighted via an estimated Kalman gain to obtain the corrected estimates of the point-valued mean and covariance of the state vector, i.e., $\bar{\bm{x}}_{i,k}$ and $\bar{P}_{i,k}$, respectively. The optimal Kalman gain denoted by $K_{j,k}~\forall j\in M_{i}$ is seen in Eq.~\eqref{eq:pDKF_measA}. For an adaptive implementation of point-valued DKF, additional parameters, namely forgetting factor~$\psi$ and pre-measurement residuals~$\bm{\epsilon}_{i,k}$, are considered in Eq.~\eqref{eq:pDKF_measB} to adaptively estimate the measurement noise covariance matrix~$R_{i,k}$ at each time iteration. 

\begin{subequations} \label{eq:pDKF_meas}
\begin{align} \label{eq:pDKF_measA}
\begin{split}
\bar{P}_{i,k}^{-1} &= \hat{P}^{-1}_{i,k} + \sum_{j\in M_{i}}\Big(H^{\top}_{j}R_{j,k}^{-1}H_{j}\Big), \\
\bar{\bm{x}}_{i,k} &= \hat{\bm{x}}_{i,k} + \sum_{j\in M_{i}}K_{j,k}\Big(\bm{z}_{j,k}-H_{j}\hat{\bm{x}}_{i,k}\Big),
\end{split}
\end{align}
where 
\begin{align} \label{eq:pDKF_measB}
\begin{split} 
\bm{\epsilon}_{j,k} &=\bm{z}_{j,k}-H_{j}\hat{\bm{x}}_{j,k}, \\
R_{j,k} &= \psi R_{j,k-1} + (1-\psi)(\bm{\epsilon}_{j,k}\bm{\epsilon}^{\top}_{j,k}+H_{j}\hat{P}_{j,k} H^{\top}_{j}), \\
K_{j,k} &= \bar{P}_{i,k} H^{\top}_{j}R^{-1}_{j,k}. \\
\end{split}
\end{align}
\end{subequations}

\section{THE PROPOSED SR-DKF: A SET-VALUED SECURE STATE ESTIMATION ALGORITHM} \label{sec:algorithm}
Based on the design aspects discussed in Section~\ref{sec:problem}, we first outline the architecture of the SR-DKF and later describe the step-by-step algorithmic details. 

\begin{figure}[h]
	\setlength{\belowcaptionskip}{-4pt}
	\centering	\includegraphics[width=0.6\columnwidth]{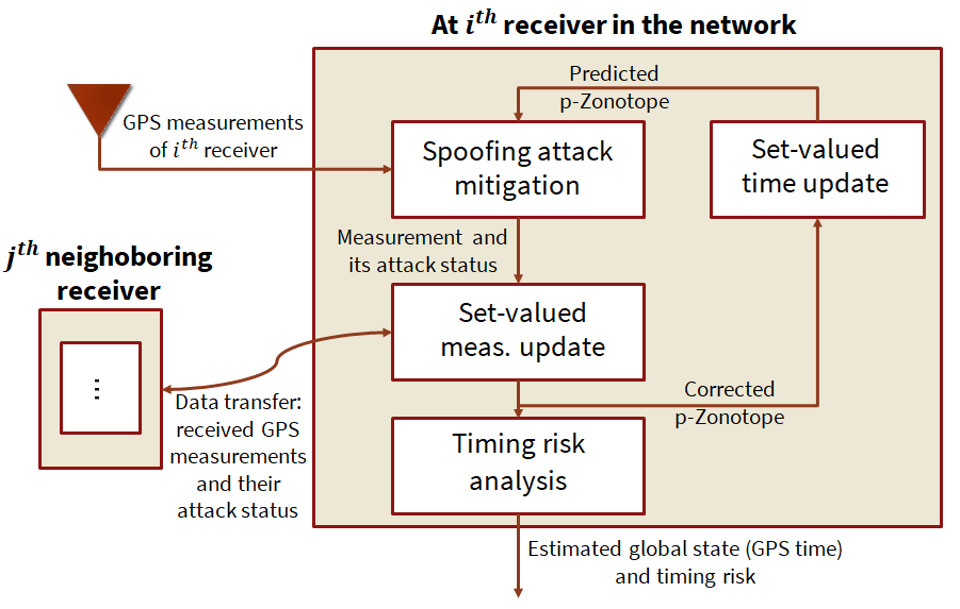}
	\caption{Architecture of the proposed SR-DKF algorithm, which is executed independently at the $i$th receiver, comprises four main modules, namely time update of set-valued DKF, spoofing attack mitigation, measurement update of set-valued DKF, and timing risk analysis.}
	\label{fig:architecture}
\end{figure}

Fig.~\ref{fig:architecture} shows the flowchart of the proposed SR-DKF algorithm that is executed independently at the $i$th receiver/power substation, and comprises four main modules, namely time update of set-valued DKF, spoofing attack mitigation, measurement update of set-valued DKF, and timing risk analysis. First, we perform a set-valued time update to compute the predicted p-Zonotope of the global state vector. Next, at each $i$th receiver, we independently utilize the predicted p-Zonotope and received GPS measurements to adaptively estimate its attack status. Thereafter, the GPS measurements and their estimated attack status are asynchronously broadcast across receivers within the network. In the measurement update, the GPS measurements and their attack status from neighbors, i.e., $j\in M_{i}$, are collectively processed to compute the corrected p-Zonotope. The center mean of the corrected p-Zonotope is the estimated point-valued GPS time that is given to PMUs. The center uncertainty and over-bounding covariance are further analyzed for their intersection probability with an unsafe set to compute the associated timing risk. In this context, we define \textit{unsafe set} by a set of global states that violate the IEEE C37.118.1a-2014 standards. Thereafter, the process repeats for the next time iteration. 
 
As explained earlier in Section~\ref{sec:intro} and from prior literature~\cite{bhamidipati2018multiple}, it is justified to consider the error bounds to be known in the non-spoofed (or authentic) conditions. We can empirically calculate the measurement bounds in authentic conditions by analyzing the large amounts of historical GPS data from publicly-accessible websites~\cite{heng2010gps}, namely Crustal Dynamics Data Information System~(CDDIS), International GNSS Service~(IGS), Continuously Operating Reference Station~(CORS), etc. Similarly, we can obtain empirical bounds on process noise by evaluating the accuracy of protocols used in power systems for synchronization across power substations~\cite{watt2015understanding} , such as Precise Time Protocol~(PTP), Network Time Protocol~(NTP), etc. 

From Eq.~\eqref{eq:pZono}, we define the p-Zonotopes of process and measurement noise bounds, which are represented by $\mathcal{L}^{i}_{\bm{\nu}}$ and $\mathcal{L}^{i}_{\bm{\omega}}$, respectively, as follows: 
\begin{align} \label{eq:noise}
\begin{aligned}
\mathcal{L}^{i}_{\bm{\nu}} &=\big(\textbf{0},\, G^{i}_{\bm{\nu}}, \,\Sigma^{i}_{\bm{\nu}}\big) \\
\mathcal{L}^{i}_{\bm{\omega}} &=\big(\textbf{0},\, G^{i}_{\bm{\omega}}, \,\Sigma^{i}_{\bm{\omega}}\big)
\end{aligned}
\end{align}
\noindent where $G^{i}_{\bm{\nu}}$ and $\Sigma^{i}_{\bm{\nu}}$ denote the generator and covariance matrices of p-Zonotope that represents process noise bounds. Similarly, $G^{i}_{\bm{\omega}}$ and $\Sigma^{i}_{\bm{\omega}}$ denote the generator and covariance matrices of p-Zonotope that represents measurement noise bounds. Note that the over-bounding covariance matrices $\Sigma^{i}_{\bm{\nu}}$ and $\Sigma^{i}_{\bm{\omega}}$ represent the characteristics of p-Zonotopes and are different from $Q_{i,k}$ and $R_{i,k}$, defined earlier in Section~\ref{sec:pntDKF}.   

\subsection{Set-valued DKF} \label{sec:setDKF}
We formulate the set-valued DKF by applying stochastic reachability to the point-valued DKF described earlier in \newline Eqs.~\eqref{eq:pDKF_init}-\eqref{eq:pDKF_time}. Similar to point-valued DKF, the set-valued DKF also performs time and measurement updates to compute the predicted and corrected stochastic reachable sets of the global state vector~$\bm{x}_{k}$, respectively. To perform spoofing attack mitigation and timing risk analysis across a network of GPS receivers, we analyze the predicted and corrected stochastic reachable sets of the state estimation error, respectively. At the $k$th iteration, the estimation error in the point-valued predicted and corrected state estimate is given by~$\Delta\hat{\bm{x}}_{i,k}$ and $\Delta\bar{\bm{x}}_{i,k}$, respectively, where~$\Delta\hat{\bm{x}}_{i,k}=\hat{\bm{x}}_{i,k}-\bm{x}_{k}$ and $\Delta\bar{\bm{x}}_{i,k}=\bar{\bm{x}}_{i,k}-\bm{x}_{k}$. 

We represent the predicted and corrected stochastic reachable sets of state estimation error via predicted and corrected p-Zonotopes, respectively. The predicted and corrected p-Zonotopes of the state estimation error denoted by $\mathcal{L}^{i}_{\Delta\hat{\bm{x}},k}$ and $\mathcal{L}^{i}_{\Delta\bar{\bm{x}},k}$, respectively, are given by
\begin{align} \label{eq:pZonoSymbol}
\begin{aligned}
\mathcal{L}^{i}_{\Delta\bar{\bm{x}},k} &= \big(\Delta\bar{\bm{x}}_{k},\, G^{i}_{\Delta\bar{\bm{x}},k}, \,\Sigma^{i}_{\Delta\bar{\bm{x}},k}\big), \\
\mathcal{L}^{i}_{\Delta\hat{\bm{x}},k} &=\big(\Delta\hat{\bm{x}}_{k},\, G^{i}_{\Delta\hat{\bm{x}},k}, \,\Sigma^{i}_{\Delta\hat{\bm{x}},k}\big),
\end{aligned}
\end{align}
\noindent where $G^{i}_{\Delta\bar{\bm{x}}}$ and $\Sigma^{i}_{\Delta\bar{\bm{x}}}$ denote the generator and covariance matrices of the corrected p-Zonotope of the state estimation error. The corrected p-Zonotope is estimated during the measurement update of set-valued DKF, which is explained later in Section~\ref{sec:pZonomeas}. Similarly, $G^{i}_{\Delta\hat{\bm{x}}}$ and $\Sigma^{i}_{\Delta\hat{\bm{x}}}$ denote the generator and covariance matrices of the predicted p-Zonotope of the state estimation error. This predicted p-Zonotope is estimated during the time update of set-valued DKF, which is explained later in Section~\ref{sec:pZonotime}. Note that the above-defined covariance matrices $\Sigma^{i}_{\Delta\hat{\bm{x}}}$ and $\Sigma^{i}_{\Delta\bar{\bm{x}}}$ represent the Gaussian characteristics of the p-Zonotopes and are different from $\hat{P}_{i,k}$ and $\bar{P}_{i,k}$, defined earlier in Section~\ref{sec:pntDKF}.

At the $k$th iteration, we utilize the translation set property in Eq.~\eqref{eq:pZonoProp3} to represent the predicted and corrected p-Zonotopes of the global state vector~$\mathcal{L}^{i}_{\hat{\bm{x}},k}$ and $\mathcal{L}^{i}_{\bar{\bm{x}},k}$, respectively, as $\mathcal{L}^{i}_{\hat{\bm{x}},k}=\bm{x}_{k}+\mathcal{L}^{i}_{\Delta\hat{\bm{x}},k}$ and $\mathcal{L}^{i}_{\Delta \bar{\bm{x}},k}=\bm{x}_{k}+\mathcal{L}^{i}_{\Delta \bar{\bm{x}},k}$. Therefore, the predicted and corrected p-Zonotopes of the global state vector are given by Eq.~\eqref{eq:pZonoSymbol2}. Note that translating the center of p-Zonotope as seen in Eq.~\eqref{eq:pZonoSymbol2} does not change the generator and covariance matrix. Therefore, the generator~$G$ and covariance matrix~$\Sigma$ of the predicted (or corrected) p-Zonotope of the state estimate~$\bm{x}$ and state estimation error~$\Delta\bm{x}$ are the same. From Eq.~\eqref{eq:pZonoSymbol2} and prior literature~\cite{shi2014set,althoff2009safety}, it is established that the center mean of predicted and corrected p-Zonotopes is equivalent to their point-valued DKF estimates. 
\begin{align} \label{eq:pZonoSymbol2}
\begin{aligned}
\mathcal{L}^{i}_{\hat{\bm{x}},k} &=\bm{x}_{k}+\mathcal{L}^{i}_{\Delta\hat{\bm{x}},k}=\big(\hat{\bm{x}}_{k},\, G^{i}_{\Delta\hat{\bm{x}},k}, \,\Sigma^{i}_{\Delta\hat{\bm{x}},k}\big) \\
\mathcal{L}^{i}_{\bar{\bm{x}},k} &=\bm{x}_{k}+\mathcal{L}^{i}_{\Delta \bar{\bm{x}},k}=\big(\bar{\bm{x}}_{k},\, G^{i}_{\Delta\bar{\bm{x}},k}, \,\Sigma^{i}_{\Delta\bar{\bm{x}},k}\big)
\end{aligned}
\end{align} 

\subsubsection{Predicted p-Zonotope of the state estimation error via time update } \label{sec:pZonotime}

\begin{figure}[H]
	\centering
	\begin{subfigure}[b]{0.3\textwidth}
		\includegraphics[width=\textwidth]{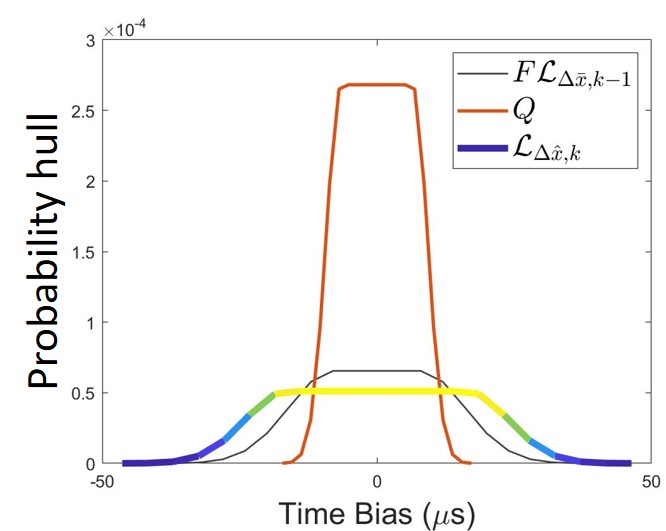}
		\caption{Set-valued time update: Time bias}
	\end{subfigure}
	\hspace{10mm} 
	\begin{subfigure}[b]{0.3\textwidth}
		\includegraphics[width=\textwidth]{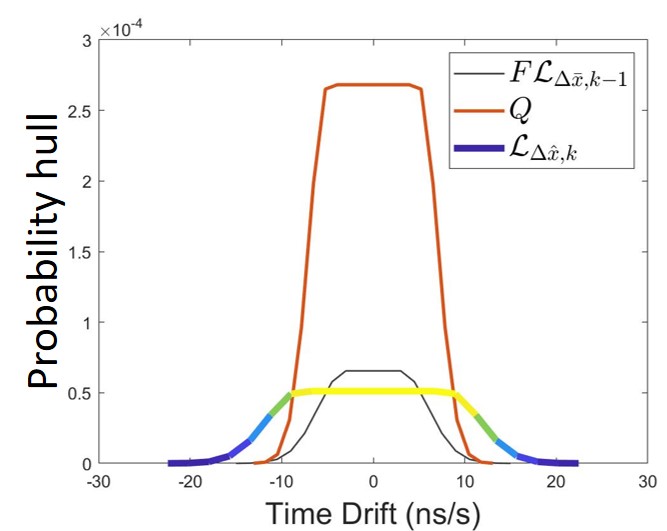}
		\caption{Set-valued time update: Time drift}
	\end{subfigure}
	\caption{Visualization of the predicted p-Zonotope of the state estimation error after the time update of set-valued DKF. The 2D p-Zonotope is sliced to be shown as two 1D p-Zonotopes, one each for (a) time bias; (b) time drift. The Minkowski sum of the linearly mapped predicted p-Zonotope at the $k$th iteration, indicated in orange, and the pre-computed process noise bound at the $(k-1)$th iteration, indicated by gray, provide an estimate of the predicted p-Zonotope for the $k$th iteration, indicated by the Parula colormap.}
	\label{fig:pZonoTimeUpdate}
\end{figure}

In the time update of set-valued DKF, as shown in Fig.~\ref{fig:pZonoTimeUpdate}, we estimate the predicted p-Zonotope of the state estimation error for the next time iteration by utilizing the corresponding corrected p-Zonotope of the current iteration and the p-Zonotope of process noise bounds. From the time update of point-valued DKF explained earlier in Eq.~\eqref{eq:pDKF_time}, we formulate the point-valued state estimation error~$\Delta \hat{\bm{x}}_{i,k}$ as
\begin{align*} 
\begin{aligned}
\Delta \hat{\bm{x}}_{i,k} &= \hat{\bm{x}}_{i,k} -\bm{x}_{k},\\
&= F\bar{\bm{x}}_{i,k-1} -(F\bm{x}_{i,k-1}+\nu_{i,k}), \\ 
&= F\Delta \bar{\bm{x}}_{i,k-1}-\nu_{i,k}. \\
\end{aligned}
\end{align*}

Similar to the measurement update, given that $\Delta \bar{\bm{x}}_{i,k-1}$, $\bm{\nu}^{i}_{k}~\forall j\in M_{i}$ are independent random variables, we apply the properties of p-Zonotopes discussed earlier in Eqs.~\eqref{eq:pZonoProp1}-\eqref{eq:pZonoProp1B} to estimate the predicted p-Zonotope as seen in Eqs.~\eqref{eq:settime}-\eqref{eq:pZonotime}. For an example shown in Fig.~\ref{fig:pZonoMeasUpdate}, the predicted p-Zonotope obtained after the Minokwski sum in Eq.~\eqref{eq:settime} is indicated by the Parula colormap. 
\begin{align} \label{eq:settime}
\mathcal{L}^{i}_{\Delta \hat{\bm{x}},k} = F\mathcal{L}^{i}_{\Delta \bar{\bm{x}},k-1} \bigoplus \mathcal{L}^{j}_{\bm{\nu,k}},
\end{align}
such that
\begin{align} \label{eq:pZonotime}
\begin{aligned}
G^{i}_{\Delta \hat{\bm{x}},k} &= \bigg[FG^{i}_{\Delta \bar{\bm{x}},k-1},\, G^{i}_{\bm{\nu},k}\bigg], \\
\Sigma^{i}_{\Delta \hat{\bm{x}},k} &=  F\Sigma^{i}_{\Delta \bar{\bm{x}},k-1}F^{\top}+\Sigma^{i}_{\bm{\nu},k}.
\end{aligned}
\end{align}

\subsubsection{Corrected p-Zonotope of the state estimation error via measurement update } \label{sec:pZonomeas}

\begin{figure}[H]
	\centering
	\begin{subfigure}[b]{0.45\textwidth}
		\includegraphics[width=\textwidth]{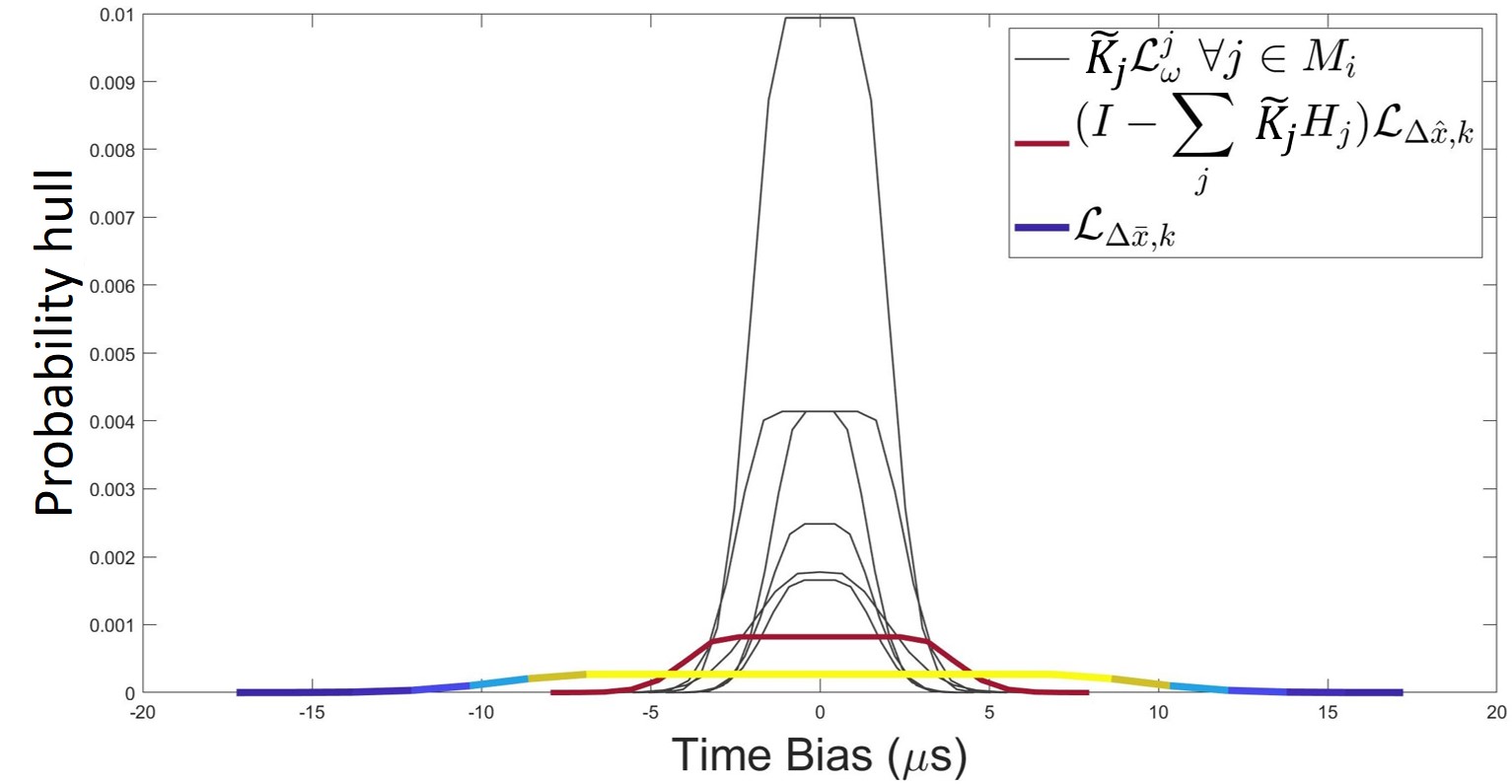}
		\caption{Set-valued measurement update: Time bias}
	\end{subfigure}
	\hspace{10mm} 
	\begin{subfigure}[b]{0.45\textwidth}
		\includegraphics[width=\textwidth]{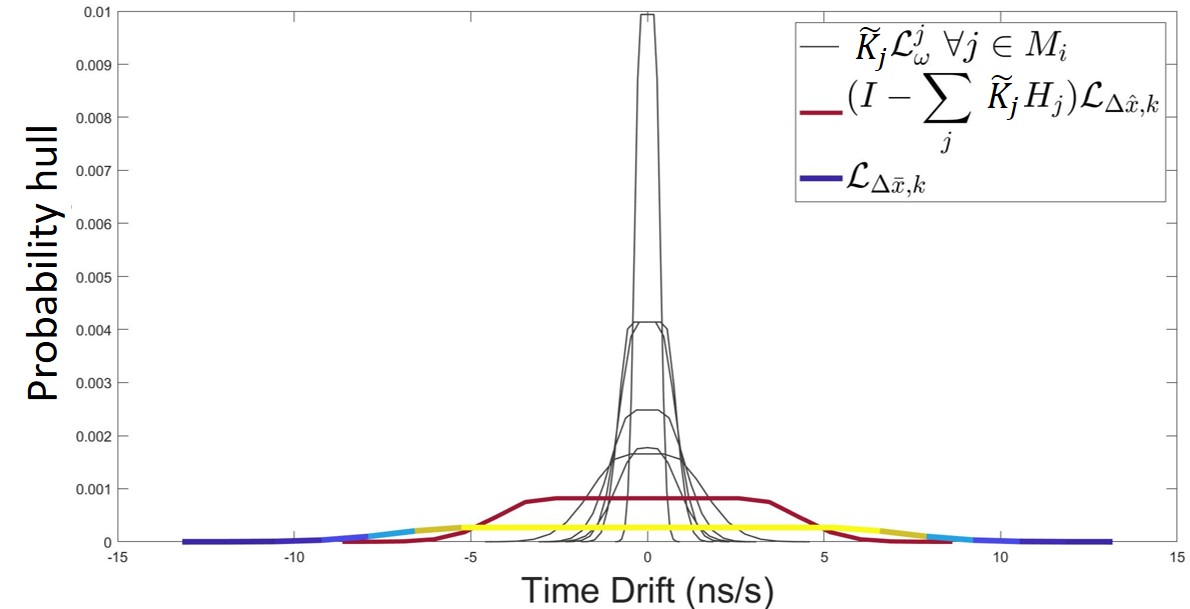}
		\caption{Set-valued measurement update: Time drift}
	\end{subfigure}
	\caption{Visualization of the corrected p-Zonotope of the state estimation error after the measurement update of set-valued DKF. The 2D p-Zonotope is sliced to be shown as two 1D p-Zonotopes, one each for (a) time bias; (b) time drift. The Minkowski sum of the linearly mapped predicted p-Zonotope at the $k$th iteration, indicated in red, and the pre-computed measurement noise bounds, indicated by gray, provide an estimate of the corrected p-Zonotope for the $k$th iteration, indicated by the Parula colormap.}
	\label{fig:pZonoMeasUpdate}
\end{figure}

In the measurement update step of set-valued DKF, as shown in Fig.~\ref{fig:pZonoMeasUpdate}, we utilize the p-Zonotope of measurement noise bounds and predicted p-Zonotope of the state estimation error to compute the corrected p-Zonotope. Unlike the conventional point-valued DKF that utilizes the optimal Kalman gain~$K_{j,k}$, we utilize an adaptive gain matrix~$\tilde{K}_{j,k}$ to perform the measurement update of the set-valued DKF. This adaptive gain matrix is estimated during the spoofing attack mitigation module, which is explained later in Section~\ref{sec:fault}. From the measurement update of point-valued DKF explained earlier in Eq.~\eqref{eq:pDKF_meas}, we formulate the point-valued state estimation error~$\Delta \bar{\bm{x}}_{i,k}$ as
\begin{align*}
\begin{aligned}
\Delta \bar{\bm{x}}_{i,k} &= \bar{\bm{x}}_{i,k} -\bm{x}_{k}\\
&= \hat{\bm{x}}_{i,k} + \sum_{j\in M_{i}}\tilde{K}_{j,k}\Big(\bm{y}_{j,k}-H_{j}\hat{\bm{x}}_{i,k}\Big) -\bm{x}_{k} \\
&= \Delta \hat{\bm{x}}_{i,k} + \sum_{j\in M_{i}}\tilde{K}_{j,k}\Big(H_{j}\bm{x}_{k}+\bm{\omega}_{j,k}-H_{j}\hat{\bm{x}}_{i,k}\Big) \\
\Delta \bar{\bm{x}}_{i,k}&= \Big(\bm{I}-\sum_{j\in M_{i}}\tilde{K}_{j,k}H_{j}\Big)\Delta \hat{\bm{x}}_{i,k} + \sum_{j\in M_{i}}\tilde{K}_{j,k}\bm{\omega}_{j,k}
\end{aligned}
\end{align*}

Given that $\Delta \hat{\bm{x}}_{i,k}$, $\bm{\omega}_{j,k}~\forall j\in M_{i}$ are independent random variables, we apply the properties of p-Zonotopes discussed earlier in Eqs.~\eqref{eq:pZonoProp1}-\eqref{eq:pZonoProp1B} to convert the point-valued representation of Eq.~\eqref{eq:pDKF_meas} to set-valued stochastic reachable sets as seen in Eq.~\eqref{eq:setmeas}. For an example shown in Fig.~\ref{fig:pZonoMeasUpdate}, the corrected p-Zonotope obtained after the Minokwski sum in Eq.~\eqref{eq:setmeas} is indicated by a Parula colormap. 
\begin{align} \label{eq:setmeas}
\mathcal{L}^{i}_{\Delta \bar{\bm{x}},k} = \bigg(\bm{I}-\sum_{j\in\Omega^{i}}\tilde{K}_{j,k}H_{j}\bigg) \mathcal{L}^{i}_{\Delta \hat{\bm{x}},k} \bigoplus_{j\in M_{i}} \tilde{K}_{j,k}\mathcal{L}^{j}_{\bm{\omega}}
\end{align}

This is simplified further using Eq.~\eqref{eq:pZonoProp2} to obtain Eq.~\eqref{eq:pZonomeas}. Intuitively, the estimated generator and covariance matrix, i.e., $G^{i}_{\Delta \bar{\bm{x}},k}$ and $\Sigma^{i}_{\Delta \bar{\bm{x}},k}$ provide bounds on the unknown state estimation error~$\Delta\bar{\bm{x}}_{i,k}$ associated with the estimated point-valued state estimate~$\bar{\bm{x}}_{i,k}$. The corrected p-Zonotope~$\mathcal{L}^{i}_{\Delta \bar{\bm{x}},k}$ of the state estimation error is analyzed to perform timing risk analysis in Section~\ref{sec:risk}.
\begin{align} \label{eq:pZonomeas}
\begin{aligned}
G^{i}_{\Delta \bar{\bm{x}},k} &= \Bigg[\Big(\bm{I}-\sum_{j\in M_{i}}\tilde{K}_{j,k}H_{j}\Big)G^{i}_{\Delta \hat{\bm{x}},k},\, \tilde{K}_{1,k}G^{1}_{\bm{\omega},k},\cdots,\,\tilde{K}_{\big|M_{i}\big|,k}G^{\big|M_{i}\big|}_{\bm{\omega},k}\Bigg] \\
\Sigma^{i}_{\Delta \bar{\bm{x}},k} &=  \Big(\bm{I}-\sum_{j\in M_{i}}\tilde{K}_{j,k}H_{j}\Big)\Sigma^{i}_{\Delta  \hat{\bm{x}},k}\Big(\bm{I}-\sum_{j\in M_{i}}\tilde{K}_{j,k}H_{j}\Big)^{\top}+\sum_{j\in M_{i}}\tilde{K}_{j,k}\Sigma^{j}_{\bm{\omega}}\tilde{K}^{\top}_{j,k}
\end{aligned}
\end{align}

\subsection{Spoofing attack mitigation and data transfer} \label{sec:fault}
Before measurement update at the $i$th receiver and $k$th iteration, we independently utilize the predicted p-Zonotope of the state estimation error~$\mathcal{L}^{i}_{\Delta \bm{x},k}$ from Eq.~\eqref{eq:pZonotime} and the pre-computed measurement noise bound~$\mathcal{L}^{i}_{\omega,k}$ to estimate the p-Zonotope of expected measurement innovation. We apply the set properties of p-Zonotopes in Eqs.~\eqref{eq:pZonoProp1}-\eqref{eq:pZonoProp2} on the point-valued measurement innovation $\bm{\epsilon}_{i,k} = \bm{z}_{i,k}-H_{i}\hat{\bm{x}}_{i,k}$, which is defined earlier in Eq.~\eqref{eq:pDKF_meas}. This p-Zonotope denoted by~$\mathcal{L}^{i}_{\bm{\epsilon},k}$ is given by
\begin{align} \label{eq:pZonores}
\mathcal{L}^{i}_{\bm{\epsilon},k} = \mathcal{L}^{i}_{\bm{\omega},k}\bigoplus H_{i}\mathcal{L}^{i}_{\Delta \hat{\bm{x}},k}. 
\end{align}

The p-Zonotope of expected measurement innovation~$\mathcal{L}^{i}_{\bm{\epsilon},k}$ provides a probability value corresponding to each point-valued measurement innovation~$\bm{\epsilon}_{i,k}$ and this probability is denoted by~$\alpha_{i,k}$. Based on this, for any $i$th receiver, we independently compute the attack status~$\tilde{\alpha}_{i,k}$ of the received GPS measurements by normalizing the obtained probability value~$\alpha_{i,k}$ with the probability value corresponding to the center mean of the p-Zonotope~$\mathcal{L}^{i}_{\bm{\epsilon},k}$ and subtracting it from one. The measurement attack status~$\tilde{\alpha}_{i,k}\in \mathbb{R}$ lies between $[0,1]$. As seen earlier in Eq.~\eqref{eq:noise}, the pre-computed process and measurement noise bounds are defined for non-spoofed conditions. Therefore, in a non-spoofed condition, the point-valued measurement innovation~$\bm{\epsilon}_{i,k}$ and its p-Zonotope~$\mathcal{L}^{i}_{\bm{\epsilon},k}$ of expected measurement innovation are in close agreement, and hence a low attack status~$\tilde{\alpha}_{i,k}\approx 0$ is obtained. However, during spoofing, the point-valued measurement innovation does not comply with this estimated p-Zonotope, and therefore, a high value of~$\tilde{\alpha}_{i,k}\approx 1$ is observed. 

Thereafter, each $i$th receiver in the network~$i\in M$, broadcasts its received GPS measurement~$\bm{z}_{i,k}$ and the estimated attack status~$\tilde{\alpha}_{i,k}$ to all the receivers in its neighborhood set~$j\in M_{i}-i$. The set~$M_{i}-i$ simply implies that the $i$th receiver need not broadcast GPS measurements to itself. Similarly, each $i$th receiver also receives the GPS measurements~$\bm{z}_{j,k}$ and their associated attack status~$\tilde{\alpha}_{i,k}$ from one or more receivers in its neighborhood set~$j\in M_{i}-i$. In an intuitive sense, the measurement attack status obtained from each $j$th receiver in the neighborhood set indicates its belief in the received GPS measurement~$\bm{z}_{j,k}$. Thereafter, the measurement attack status is used to compute the adaptive gain matrix~$\tilde{K}_{j,k}$ as seen in Eq.~\eqref{eq:setGain}. Given that the gain matrix is a value between $0$ and $1$, the Eq.~\eqref{eq:setGain} empirically represents a joint minimization of both spoofing probability and covariance of the point-valued state estimate. The formulated adaptive gain matrix~$\tilde{K}_{j,k}$ is later utilized to perform the set-valued measurement update of SR-DKF, which was described earlier in Eq.~\eqref{eq:pZonomeas}. 
\begin{align} \label{eq:setGain}
\tilde{K}^{i}_{j,k} &=\big(1-\tilde{\alpha}^{i}_{j,k}\big)H_{j}^{\top}\bar{P}_{i,k}R^{-1}_{j,k}~~~\forall j\in M_{i}
\end{align}  

\subsection{Timing risk analysis} \label{sec:risk}
\begin{figure}[H]
	\centering
	\begin{subfigure}[b]{0.45\textwidth}
		\includegraphics[width=\textwidth]{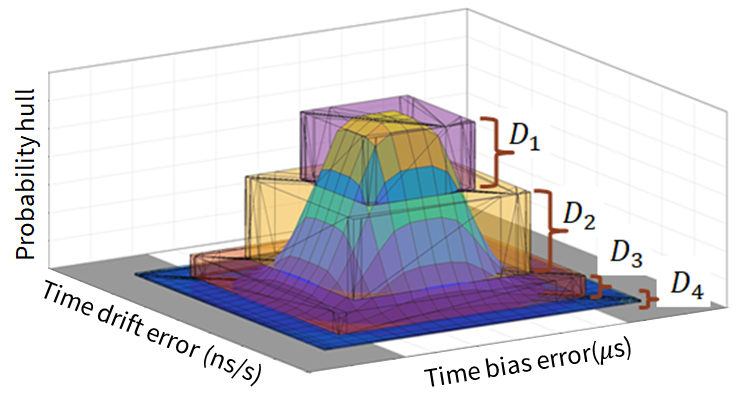}
		\caption{Over-approximation of corrected p-Zonotope}
		\label{fig:SRDKFIntRisk}
	\end{subfigure}
	\hspace{10mm} 
	\begin{subfigure}[b]{0.45\textwidth}
		\includegraphics[width=\textwidth]{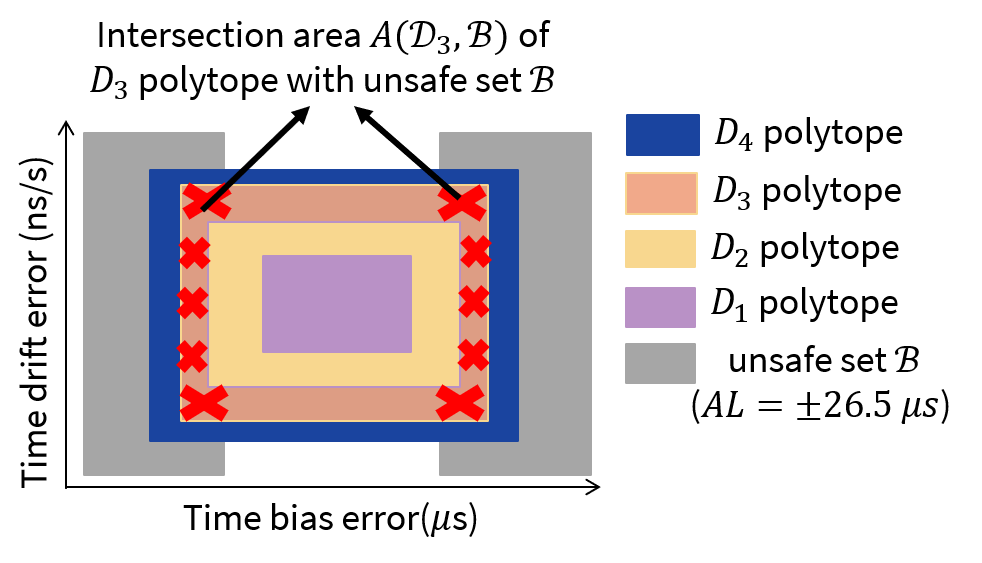}
		\caption{Top view of finite polytopes}
		\label{fig:SRDKFIntRiskTopView}
	\end{subfigure}
	\caption{Visualization of the corrected p-Zonotope~$\mathcal{L}^{i}_{\Delta \bar{x},k}$ of state estimation error and its intersection with the unsafe set~$\mathcal{B}$; (a)~shows an example that over-approximates the corrected p-Zonotope by four polytopes, which are denoted by~$D_{1}$ to $D_{4}$; (b)~shows the top-view of the finite polytopes and an example intersection region of $D_{3}$ polytope with unsafe set~$\mathcal{B}$, which is indicated by red cross marks.}
\end{figure}

We evaluate the probability of corrected p-Zonotope~$\mathcal{L}^{i}_{\Delta \bar{x},k}$ of state estimation error derived in Eq.~\eqref{eq:pZonomeas}, to intersect an unsafe set, i.e., a set of states that violate the IEEE C37.118.1a-2014 standards. As shown in both Figs.~\ref{fig:SRDKFIntRisk} and~\ref{fig:SRDKFIntRiskTopView}, the unsafe set represents two gray strips, i.e., $\Delta\bar{T}_{i,k}\in [26.5\mu s, \infty]$ and $\Delta\bar{T}_{i,k}\in [-\infty,-26.5\mu s]$. Intuitively, this matches the definition of risk described earlier in Section~\ref{sec:intro}, which denotes the probability of the state estimation error to exceed a safety AL. 

To perform the timing risk analysis, we over-approximate the corrected p-Zonotope~$\mathcal{L}^{i}_{\Delta \bar{x},k}$ by a finite series of polytopes~\cite{althoff2009safety}. An example of over-approximation of p-Zonotope by four polytopes $D_{1}$ to $D_{4}$ is shown in Fig.~\ref{fig:SRDKFIntRisk}. Thereafter, we compute the intersection area of each $l$th polytope~$D_{l}$ with the unsafe set~$\mathcal{B}$. Figure~\ref{fig:SRDKFIntRiskTopView} shows the top-view of over-approximated series of polytopes and an example intersection region of $D_{3}$ polytope with unsafe set~$\mathcal{B}$. 

The timing risk, seen in Eq.~\eqref{eq:intRisk}, is defined as the sum across polytopes~$D_{l}$ of their intersection area, which is denoted by~$\mathcal{A}(\cdot)$, with the unsafe set~$\mathcal{B}$ times the max probability of $D_{l}$th polytope, which is denoted by $p_{max,l}$. The max probability of $D_{l}$th polytope is at its top face and equals the probability of corrected p-Zonotope at that level. For computational tractability, we represent the p-Zonotope by a finite number of polytopes and a $\gamma$-confidence set. The $\gamma$-confidence set~\cite{althoff2009safety} thresholds the tail of a p-Zonotope by a pre-defined parameter~$\gamma$, such that $P[-\gamma<\mathcal{N}(0,1)<\gamma]=\textrm{erf}(\dfrac{\gamma}{\sqrt{2}})$, where $\textrm{erf}$ denotes the error function~\cite{oldham2008error}. From~\cite{althoff2009safety}, the probability that the state estimation error lies outside the $\gamma$-confidence set of corrected p-Zonotope is $1-\textrm{erf}\Big(\dfrac{\gamma}{\sqrt(2)}\Big)^{2n}$. 
\begin{align} \label{eq:intRisk}
r_{i,k} = 1-\textrm{erf}\Big(\dfrac{\gamma}{\sqrt{2}}\Big)^{2n}+\sum_{l}A(\mathcal{D}_{l}\cap\mathcal{B})\times p_{max,l}
\end{align} 

\subsection{Implementation details}
\begin{figure}[h]
	\setlength{\belowcaptionskip}{-4pt}
	\centering	\includegraphics[width=0.6\columnwidth]{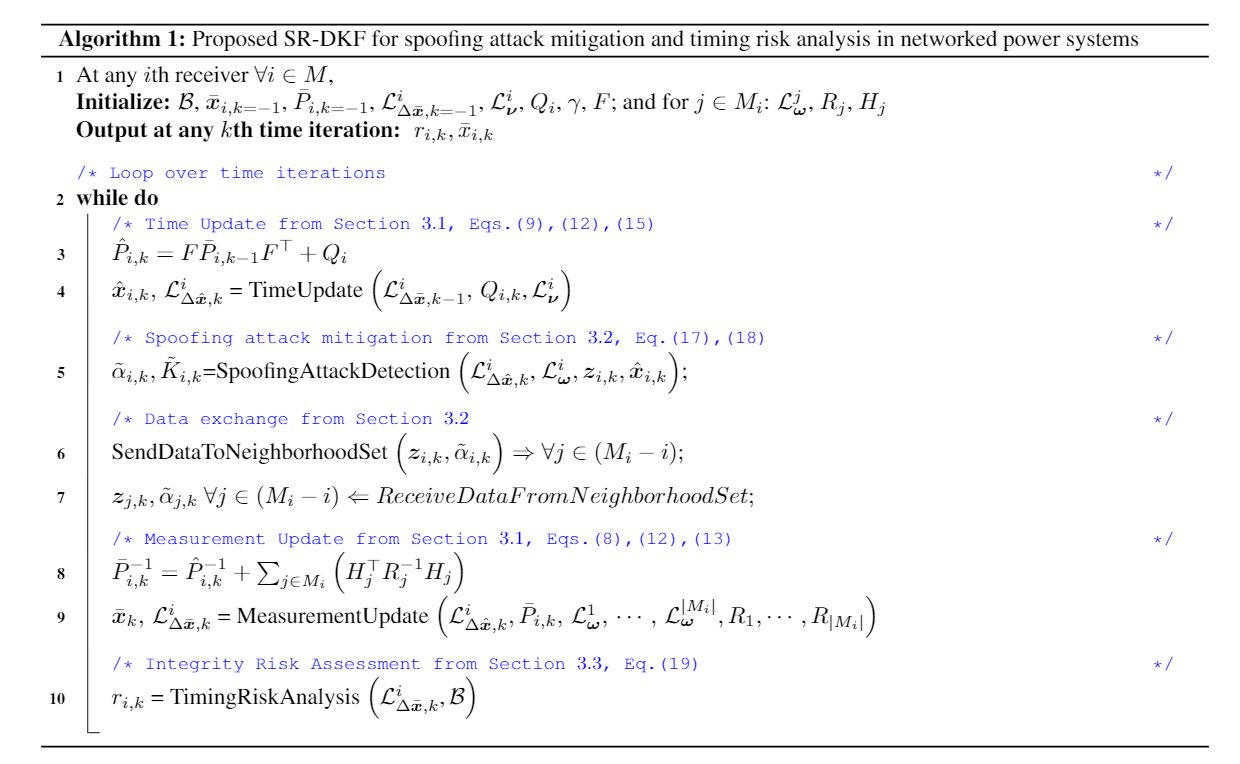}
	\caption{Implementation steps of the proposed SR-DKF algorithm.}
	\label{fig:AlgoSummary}
\end{figure}

Figure~\ref{fig:AlgoSummary} summarizes the implementation steps of the proposed SR-DKF described in Sections~\ref{sec:setDKF}-\ref{sec:risk}. We assume that the $i$th receiver~$~\forall i\in M$ is authentic during initialization. Therefore, at iteration~$k=0$, we define the predicted p-Zonotope~$\mathcal{L}^{i}_{\Delta \bar{\bm{x}},-1}$ of the state estimation error to be zero-mean center, with non-zero generator and over-bounding covariance matrices. 

\section{RESULTS AND ANALYSIS} \label{sec:results}
We validate the robustness of the proposed SR-DKF algorithm to not only mitigate the effect of spoofing but also estimate the secure GPS timing and its associated timing risk across all the receivers in the network. We perform set operations of stochastic reachability and transform across various set representations, such as polytopes, zonotopes, p-Zonotopes, etc., using a publicly-available MATLAB toolbox, known as COntinuous Reachability Analyzer (CORA)~\cite{althoff2016cora}. 

\begin{figure}[h]
	\setlength{\belowcaptionskip}{-4pt}
	\centering	\includegraphics[width=0.6\columnwidth]{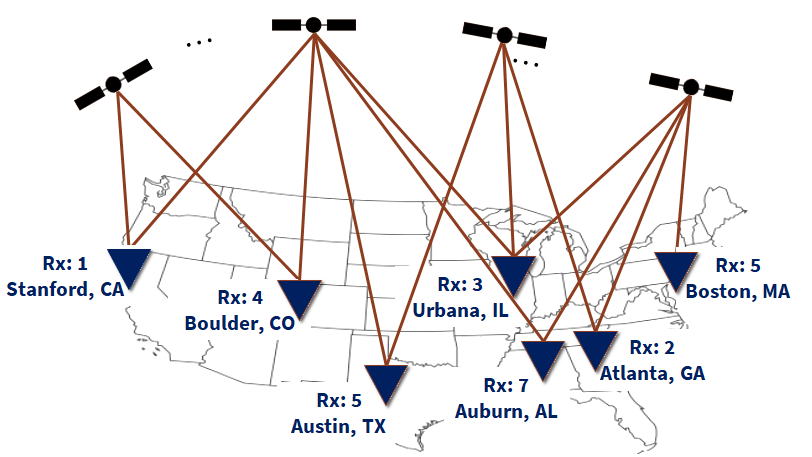}
	\caption{Geographical map showcasing a simulated network of seven spatially distributed GPS receivers across the US.}
	\label{fig:exp_setup}
\end{figure}

As shown in Fig.~\ref{fig:exp_setup}, we design a simulated network of seven GPS receivers,~$\big|M\big|=7$, that are spatially distributed across the US. The network of GPS receivers indexed from $Rx:1$ to $Rx:7$ are located in 1) Stanford, CA;  2) Atlanta, GA; 3) Urbana, IL; 4) Boulder, CO; 5) Austin, TX; 6) Boston, MA; and 7) Auburn, AL. We utilized a C++ language-based software-defined GPS simulator known as GPS-SIM-SDR~\cite{gpssdrsim,bhamidipati2019gps} to simulate the raw GPS signals received at each location in the network. Later, we post-processed the simulated GPS signals using a MATLAB-based software-defined radio known as SoftGNSS~\cite{softGNSS}. 
We define the simulated errors observed in the system during authentic conditions to have the following characteristics: 
\begin{enumerate}
	\item For the process noise in GPS time (the first global state~$T_{i,k}$), the mean and covariance of the time-varying Gaussian distribution lies between [$-2.5~\mu$s,\,$2.5~\mu$s] and [$0~\mu$s,\,$4~\mu$s], respectively; Similarly, for the process noise in its drift rate (the second global state~$\dot{T}_{i,k}$), the time-varying mean and covariance lies between [$-3.5$~ns/s,\,$3.5$~ns/s] and [$0$~ns/s,\,$6$~ns/s], respectively; 	
	\item For the measurement noise in GPS pseudoranges, the mean and covariance of the time-varying Gaussian distribution lies between [$-1~\mu$s,\,$1~\mu$s] and [$0~\mu$s,\,$3~\mu$s], respectively; Similarly, for the measurement noise in GPS Doppler, the time-varying mean and covariance lies between [$-2.5$~ns/s,\,$2.5$~ns/s] and [$0$~ns/s,\,$6$~ns/s], respectively. We consider all the receivers in this simulated setup to have the same bounds.  
	\item For the initial state error bounds in GPS time (the first global state~$T_{i,k}$), the mean and covariance of the time-varying Gaussian distribution lies between [$-1.5~\mu$s,\,$1.5~\mu$s] and [$0~\mu$s,\,$2~\mu$s], respectively; Similarly, for the initial state error bounds in the second global state~$\dot{T}_{i,k}$, the time-varying mean and covariance lies between [$-2.5$~ns/s,\,$2.5$~ns/s] and [$0$~ns/s,\,$4$~ns/s], respectively;  
\end{enumerate}

As discussed earlier in Fig.~\ref{fig:AlgoSummary}, we initialize our SR-DKF algorithm by pre-defining the following p-Zonotopes~$\forall i\in M$: process noise~$\mathcal{L}^{i}_{\bm{\nu}}$, measurement noise~$\mathcal{L}^{i}_{\bm{\omega}}$, and initial state estimation error~$\mathcal{L}^{i}_{\Delta \bar{\bm{x}},-1}$. We formulate the p-Zonotopes from the above-listed errors bounds in the same manner as the example discussed earlier in Section~\ref{sec:pZonotopes}. 

\subsection{During a coordinated signal-level spoofing attack}

\begin{figure}[H]
	\centering
	\begin{subfigure}[b]{0.3\textwidth}
		\includegraphics[width=\textwidth]{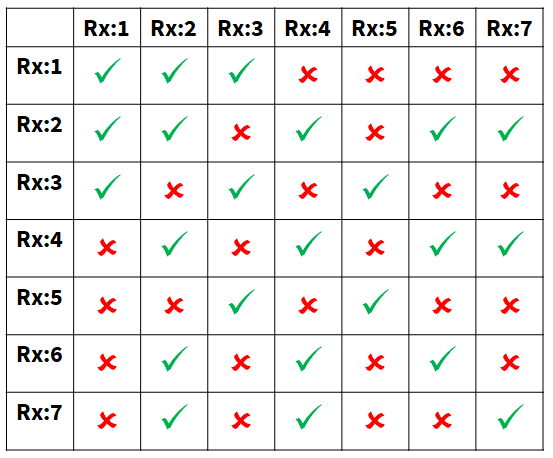}
		\caption{Sparse connectivity across the network}
		\label{fig:netmat}
	\end{subfigure}
	\begin{subfigure}[b]{0.48\textwidth}
		\includegraphics[width=\textwidth]{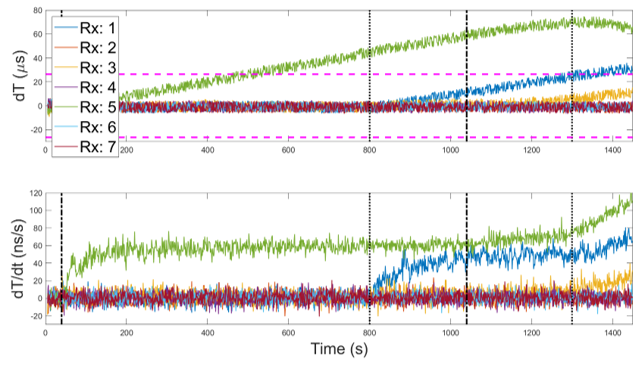}
		\caption{Point-valued state estimation via adaptive DKF}
		\label{fig:NetDKFstest}
	\end{subfigure}
	\begin{subfigure}[b]{0.48\textwidth}
		\includegraphics[width=\textwidth]{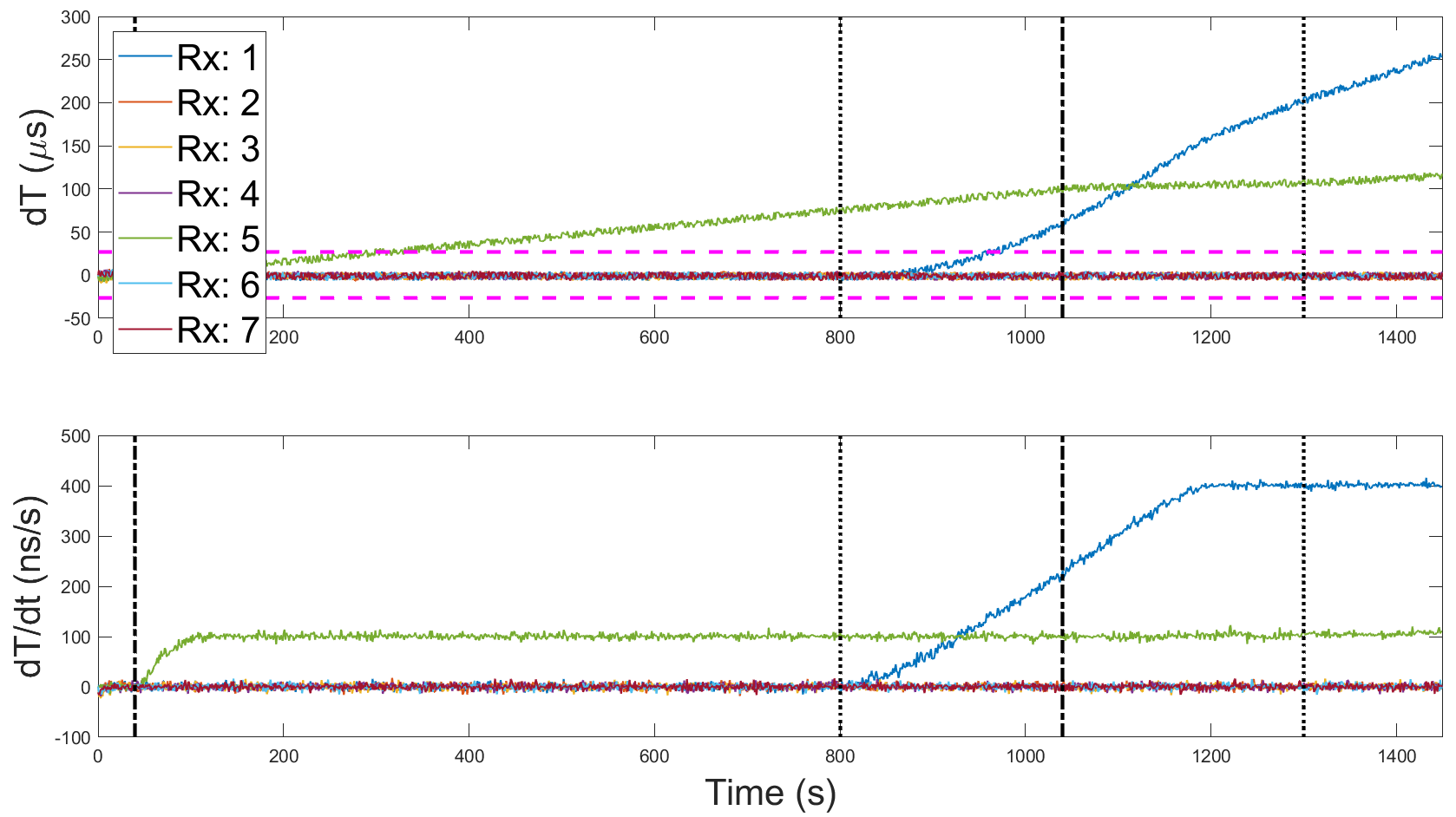}
		\caption{Point-valued adaptive KF}
		\label{fig:SingleadptKFstest}
	\end{subfigure}
	\begin{subfigure}[b]{0.48\textwidth}
		\includegraphics[width=\textwidth]{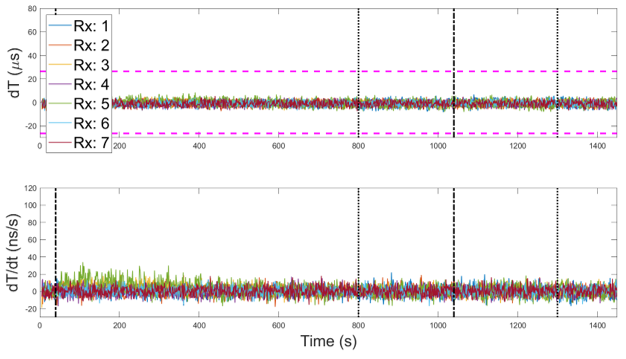}
		\caption{Set-valued state estimation via SR-DKF}
		\label{fig:SRDKFstest}
	\end{subfigure}
	\caption{Performance comparison of the three state estimation techniques; a)~sparse connectivity across the network of receivers; b)~set-valued state estimation using the proposed SR-DKF algorthm; c)~point-valued adaptive DKF; d)~point-valued adaptive KF for a single receiver; In (b)-(d) the vertical dotted black lines indicate the start and stop times of spoofing attacks while the horizontal magenta dotted lines indicate the threshold set by the IEEE C37.118.1a-2014 standards.}
\end{figure}

In the first set of experiments, we consider sparse connectivity among the seven receivers in the network. Figure~\ref{fig:netmat} showcases the connectivity matrix where a green tick indicates the presence of a bidirectional communication link while a red cross indicates otherwise. Note that each receiver is connected to itself by default. We induce a simulated coordinated signal-level spoofing attack in the simulated GPS measurements received at two GPS receivers in the network, i.e., $Rx:1$, which is located in Stanford, CA and $Rx:5$, which is located in Austin, TX. Between the time duration $k=40-1040~$s, the first spoofing attack manipulates the GPS time at $Rx:5$ to deviate at a rate of $100~$ns/s, and the second spoofing attack, which occurs between $k=800-1300~$s, deviates the GPS time of $Rx:1$ at a rate of $400~$ns/s. 

In Fig.~\ref{fig:NetDKFstest}, we show the performance of the point-valued state estimation via an adaptive DKF framework, where even the receivers that are not directly spoofed, e.g., $Rx:3$, exhibit high state estimation errors. Also, in Fig.~\ref{fig:SingleadptKFstest}, we show the state estimation accuracy achieved via a point-valued single-receiver-based adaptive KF. In both these point-valued approaches, the measurement noise covariance~$R_{i,k}$ is adaptively estimated via Eq.~\eqref{eq:pDKF_meas} by setting the forgetting factor~$\psi=0.3$. Based on the maximum error values listed in Table~\ref{tab:my-table}, we observe that both the approaches violate the IEEE C37.118.1a-2014 standards. In Fig.~\ref{fig:SRDKFstest}, we validate that our proposed SR-DKF algorithm provides a secure estimate of GPS timing with maximum error of $8.4~\mu$s across all receivers, and thereby complies with the IEEE standards at all times. The quantitative maximum error values of the receiver network for the above-described state estimation approaches are listed in Table~\ref{tab:my-table}.  

\begin{table}[H]
	\centering
	\caption{Comparison among the maximum error values of the global state vector estimated via three approaches: the proposed SR-DKF algorithm, point-valued adaptive DKF, and point-valued single-receiver-based adaptive KF.}
	\label{tab:my-table}
	\begin{tabular}{|l|l|l|l|l|l|l|l|l|}
		\hline
		\multicolumn{2}{|l|}{State estimation approach} & $Rx: 1$ & $Rx: 2$ & $Rx:3$ & $Rx:4$ & $Rx:5$ & $Rx:6$ & $Rx:7$\\ \hline
		\multicolumn{1}{|c|}{} & \textbf{\begin{tabular}[c]{@{}l@{}}Set-valued \\ SR-DKF\end{tabular}} & \textbf{8.27} & \textbf{7.42} & \textbf{6.81} & \textbf{6.79} & \textbf{8.43} & \textbf{7.10} & \textbf{7.40}\\ \cline{2-9} 
		\multicolumn{1}{|c|}{} & \begin{tabular}[c]{@{}l@{}}Point-valued \\ adaptive DKF\end{tabular} & {\color[HTML]{FE0000} 35.88} & 6.47 & 15.24 & 6.70 & {\color[HTML]{FE0000} 74.68} & 6.26 & 6.36 \\ \cline{2-9} 
		\multicolumn{1}{|c|}{\multirow{-3}{*}{$T~(\mu$s)}} & \begin{tabular}[c]{@{}l@{}}Point-valued \\ adaptive KF\end{tabular} & {\color[HTML]{FE0000} 257.36} & 7.36 & 9.48 & 6.43 & {\color[HTML]{FE0000} 118.90} & 6.46 & 6.48 \\ \hline
		& \textbf{\begin{tabular}[c]{@{}l@{}}Set-valued \\ SR-DKF\end{tabular}} & \textbf{22.05} & \textbf{19.12} & \textbf{20.54} & \textbf{17.12} & \textbf{33.23} & \textbf{16.98} & \textbf{18.01} \\ \cline{2-9} 
		& \begin{tabular}[c]{@{}l@{}}Point-valued \\ adaptive DKF\end{tabular} & 80.68 & 23.60 & 39.94 & 19.95 & 125.23 & 18.01 & 20.88 \\ \cline{2-9} 
		\multirow{-3}{*}{$\dot{T}$~(ns/s)} & \begin{tabular}[c]{@{}l@{}}Point-valued \\ adaptive KF\end{tabular} & 415.39 & 17.90 & 22.75 & 18.71 & 122.94 & 16.21 & 17.04 \\ \hline
	\end{tabular}
\end{table}

\begin{figure}[H]
	\setlength{\belowcaptionskip}{-4pt}
	\centering	\includegraphics[width=0.65\columnwidth]{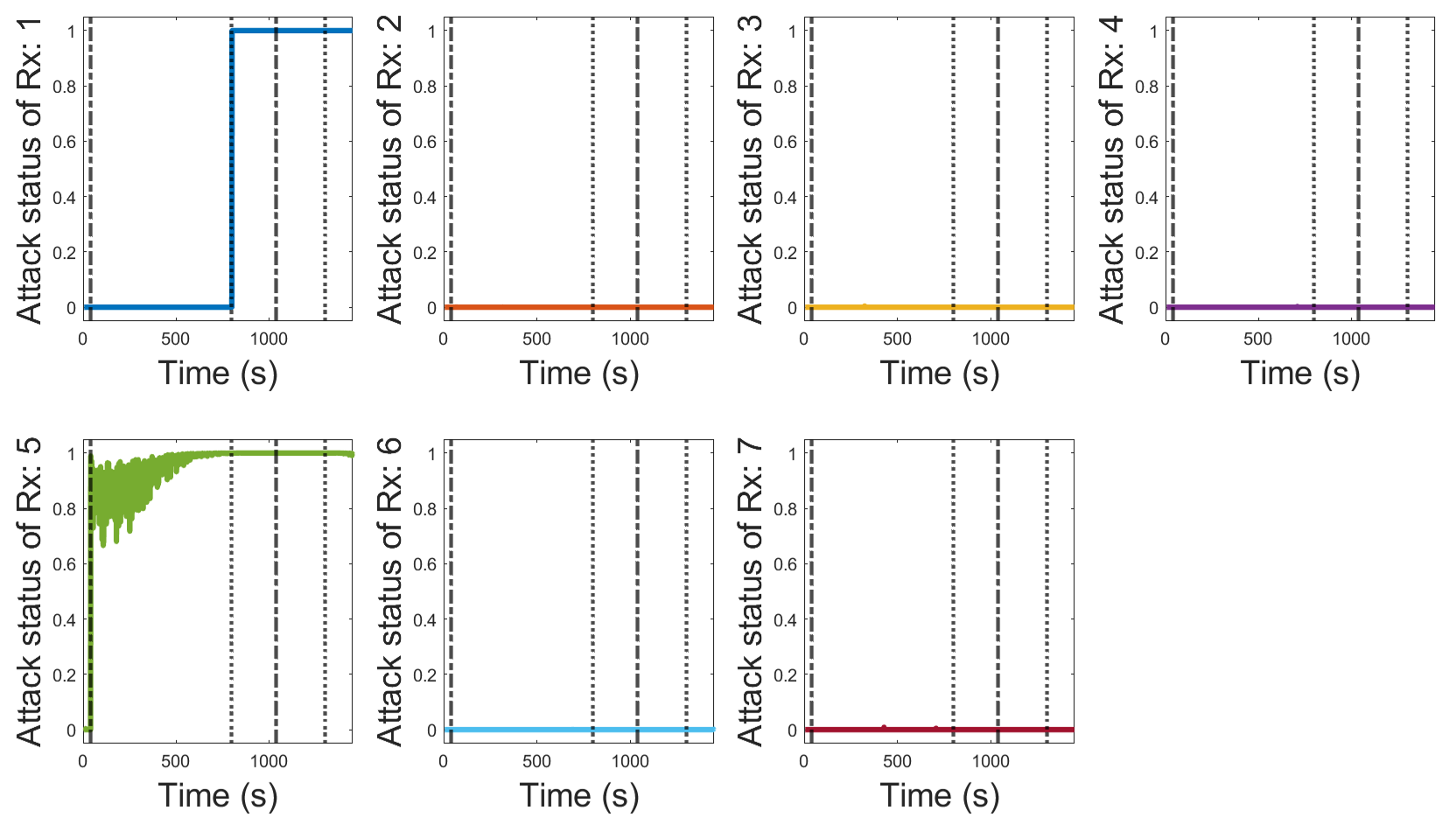}
	\caption{Attack status of each receiver in the network that is estimated via the proposed SR-DKF algorithm. This is computed by evaluating the point-valued measurement innovation against the estimated p-Zonotope of expected measurement innovation, as seen in Eq.~\eqref{eq:pZonores}. Our SR-DKF adaptively mitigates spoofing in $Rx:1,5$ and successfully estimates other receivers to be non-spoofed. }
	\label{fig:CoordSpAttackAttackStatus}
\end{figure}

In Fig.~\ref{fig:CoordSpAttackAttackStatus}, we plot the attack status~$\tilde{\alpha}_{i,k}$ of the GPS measurements~$\bm{z}_{i,k}$ at each receiver in the network. In accordance with Section~\ref{sec:fault}, we demonstrate that the proposed SR-DKF algorithm successfully detects and mitigates spoofing in the GPS measurements of $Rx:1$ and $5$ while accurately categorizing the other receivers, i.e., $Rx:2,3,4,6,7$ to be non-spoofed with attack status~$\tilde{\alpha}_{i,k}\approx 0$ at all times. We observe that the spoofing attack with higher drift rate is mitigated quicker, i.e., $400$~ns/s in $Rx:1$, as compared to the one with lower drift rate, i.e., $100$~ns/s in $Rx:5$.

\begin{figure}[H]
	\setlength{\belowcaptionskip}{-4pt}
	\centering	\includegraphics[width=0.65\columnwidth]{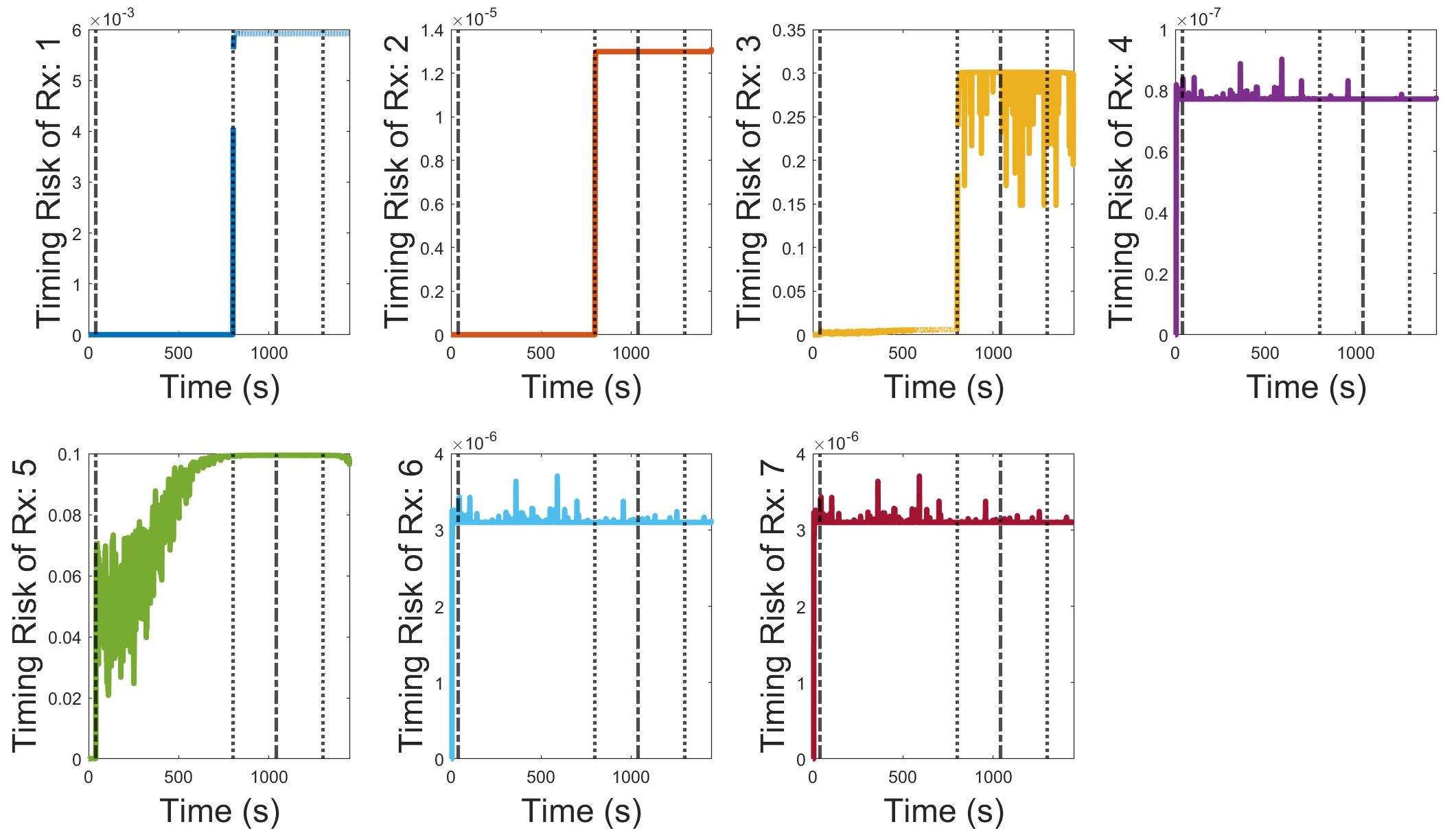}
	\caption{Robust timing risk analysis for each receiver in the network via the proposed SR-DKF algorithm. This is computed by evaluating the intersection of the corrected p-Zonotope of the state estimation error with the unsafe set as seen in Eq.~\eqref{eq:intRisk}. Our SR-DKF successfully quantifies the risk of violating the IEEE C37.118.1a-2014 standards. }
	\label{fig:CoordSpAttackIntRisk}
\end{figure}

In Fig.~\ref{fig:CoordSpAttackIntRisk}, we showcase the timing risk associated with the estimated GPS timing, i.e., the probability of GPS time to violate the IEEE C37.118.1a-2014 standards. This is computed by analyzing the intersection of the corrected p-Zonotope~$\mathcal{L}^{i}_{\Delta\bar{\bm{x}},k}$ with the unsafe set~$\mathcal{B}$, as discussed in Section~\ref{sec:risk}. An increase or jump in measure of timing risk is based on the adaptive gain matrix~$\tilde{K}_{j,k}~\forall j\in M_{i}$ that in turn depends on the estimated attack status~$\tilde{\alpha}_{j,k}$ of the receiver. Therefore, an increase in the timing risk for $Rx:5$ till $t=1040~$s is consistent with the decrease in measurement reliability, which is evident from Fig~\ref{fig:CoordSpAttackAttackStatus}. By comparing Fig.~\ref{fig:netmat} and~Fig.~\ref{fig:CoordSpAttackIntRisk}, we observe a direct correlation between the value of timing risk to the number of reliable measurements available for the calculation of corrected p-Zonotope. For instance, since none of the four neighbors of $Rx:4$ are spoofed, the estimated timing risk is consistently low, i.e., $\approx 10^{-7}$, for the entire experiment duration. In contrast, $Rx:3$ has significantly high timing risk~$(\approx 0.3)$ because two out of its three neighbors are spoofed. Similarly, when only one out of three neighbors at $Rx:1$ is spoofed, the order of timing risk magnitude is $10^{-3}$ whereas when one out of two neighbors are spoofed at $Rx:5$, the timing risk is a higher order magnitude of $0.1$. Therefore, our SR-DKF estimates a robust measure of the timing risk across all receivers in the network. 
 
\vspace{-5mm}

\subsubsection{Robustness analysis of the proposed SR-DKF algorithm}
In the second set of experiments, we demonstrate the robustness of the proposed SR-DKF algorithm as the number of receivers in the neighborhood set~$|M|$ is varied. Considering a total experiment duration of $100$~s and a network of seven receivers seen in Fig.~\ref{fig:exp_setup}, we induced a simulated meaconing attack between $k=10-100~$s in the GPS measurements that is received at $Rx:1$ located at Stanford, CA.   

\begin{figure}[H]
	\setlength{\belowcaptionskip}{-4pt}
	\centering	\includegraphics[width=0.65\columnwidth]{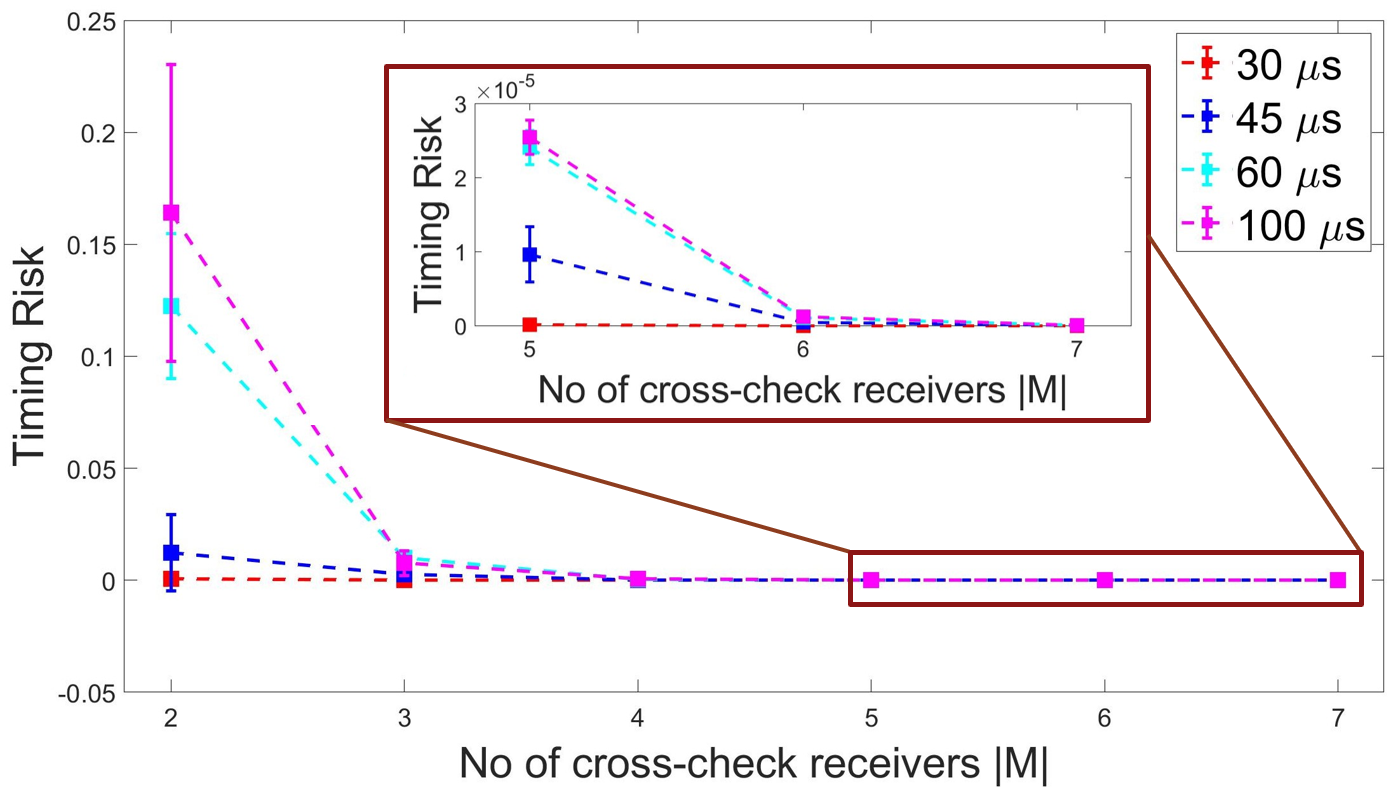}
	\caption{Variation in the timing risk associated with GPS timing of $Rx:1$ with the number of receivers in a fully-connected network. The colors red, blue, cyan and magenta indicate a meaconing attack of $30~\mu$s, $45~\mu$s, $60~\mu$s, and $100~\mu$s, respectively. We observe that the variation in timing risk decreases with an increase in the number of interacting receivers with $Rx:1$ and the magnitude of meaconing attack.}
	\label{fig:MeacVariation}
\end{figure}

In Fig.~\ref{fig:MeacVariation}, we analyzed the variation in the timing risk associated with the estimated GPS time of $Rx:1$ as the number of receivers in the network~$M$ is increased. For each size of the receiver network that ranges from $|M|=2$ to $|M|=7$, we consider the presence of a fully-connected bidirectional communication network and perform $50$ Monte-Carlo runs for four meaconing magnitudes of $30~\mu$s, $45~\mu$s, $60~\mu$s, and $100~\mu$s. We successfully demonstrated that as the number of receivers interacting with the $Rx:1$, i.e., $|M_{1}|$, increases, the associated timing risk decreases. Particularly, for $|M_{1}|>5$, the decrease in timing risk is on the order of $\leq 10^{-5}$ and hence, considered insignificant. Similarly, the effect of meaconing magnitude on the timing risk decreases as the number of interacting receivers associated with $Rx:1$ increases. Therefore, we validated the proposed SR-DKF algorithm that requires only a handful of distributed GPS receivers to achieve robust performance.   

\section{CONCLUSIONS} \label{sec:conclusion} 
In summary, we developed a Stochastic Reachability-based Distributed Kalman Filter~(SR-DKF) to perform set-valued state estimation using a network of GPS receivers that estimate not only the point-valued mean but also the stochastic reachable set of state (i.e., GPS time and its drift rate). 
By parameterizing the stochastic reachable set via probabilistic zonotopes~(p-Zonotopes), we derived the time and measurement update steps of the set-valued DKF to estimate the predicted and corrected p-Zonotope of timing error, respectively.
To compute secure GPS time for a network of receivers, we designed a two-tiered approach: first, we performed spoofing mitigation at measurement-level via deviation of measurement innovation from its expected stochastic set (represented by the p-Zonotope). Second, we performed timing risk analysis at state-level via intersection probability of corrected p-Zonotope with an unsafe set, i.e., a set that violates the IEEE C37.118.1a-2014 standards. To our knowledge, no prior literature exists on secure state estimation via stochastic reachability nor on timing risk quantification during spoofing. 

Under a simulated coordinated signal-level spoofing that affects two among a network of seven sparsely-connected GPS receivers, the proposed SR-DKF algorithm demonstrated a higher timing accuracy, successfully mitigated the spoofing attacks, and estimated a robust measure of timing risk. While our algorithm showed a maximum timing error of only $<8.4~\mu$s, both the conventional DKF and single-receiver-based adaptive KF showed high estimation errors that violate the IEEE C37.118.1a-2014 threshold of $26.5~\mu$s. Furthermore, by varying the number of receivers and attack magnitudes, we performed extensive Monte Carlo runs to validate the sensitivity of estimated timing risk.

\section*{ACKNOWLEDGEMENTS}
We would like to thank Ashwin Kanhere, Siddharth Tanwar, and Tara Yasmin Mina for their help in revising the paper and presentation. We would also like to thank the members of NAVLab research at Stanford University for their thoughtful input on this work and paper.
~\\

This material is based upon work supported by the Department of Energy under Award Number DE-OE0000780.

This report was prepared as an account of work sponsored
by an agency of the United States Government. Neither the
United States Government nor any agency thereof, nor any of
their employees, makes any warranty, express or implied, or
assumes any legal liability or responsibility for the accuracy,
completeness, or usefulness of any information, apparatus,
product, or process disclosed, or represents that its use would
not infringe privately owned rights. Reference herein to any
specific commercial product, process, or service by trade
name, trademark, manufacturer, or otherwise does not necessarily constitute or imply its endorsement, recommendation,
or favoring by the United States Government or any agency
thereof. The views and opinions of authors expressed herein
do not necessarily state or reflect those of the United States
Government or any agency thereof.

\bibliographystyle{ieeetran}
\bibliography{IEEEabrv,mybiblibrary}

\end{document}